\documentclass{aastex61}

\usepackage[latin1]{inputenc}
\usepackage{epstopdf}
\usepackage{amssymb}
\usepackage{amsmath}
\usepackage{hyperref}
\usepackage{color}

\newcommand{\nickel}{$^{56}$Ni}
\newcommand{\kms}{km\,s$^{-1}$}
\newcommand{\msolar}{\ensuremath{\mathrm{M}_{\sun}}}
\newcommand{\nm}{DES16C2nm}

\shortauthors{Smith et al.}
\shorttitle{DES16C2nm: A superluminous supernova at $z=2.0$}
     
\begin{document}

\title{Studying the Ultraviolet Spectrum of the First Spectroscopically Confirmed Supernova at redshift two}
\author{M.~Smith}
\affiliation{Department of Physics and Astronomy, University of Southampton, Southampton, SO17 1BJ, UK}

\author{M.~Sullivan}
\affiliation{Department of Physics and Astronomy, University of Southampton, Southampton, SO17 1BJ, UK}

\author{R.~C.~Nichol}
\affiliation{Institute of Cosmology and Gravitation, University of Portsmouth, Portsmouth PO1 3FX, UK}

\author{L.~Galbany}
\affiliation{PITT PACC, Department of Physics and Astronomy, University of Pittsburgh, Pittsburgh, PA 15260, USA}

\author{C.~B.~D'Andrea}
\affiliation{Department of Physics and Astronomy, University of Pennsylvania, Philadelphia, PA 19104, USA}

\author{C.~Inserra}
\affiliation{Department of Physics and Astronomy, University of Southampton, Southampton, SO17 1BJ, UK}

\author{C.~Lidman}
\affiliation{ARC Centre of Excellence for All-sky Astrophysics (CAASTRO)}
\affiliation{Australian Astronomical Observatory, North Ryde, NSW 2113, Australia}

\author{A.~Rest}
\affiliation{Space Telescope Science Institute, 3700 San Martin Drive, Baltimore, MD 21218, USA}
\affiliation{Department of Physics and Astronomy, The Johns Hopkins University, 3400 North Charles Street, Baltimore, MD 21218, USA}

\author{M.~Schirmer}
\affiliation{Gemini Observatory, Casilla 603, La Serena, Chile}

\author{A.~V.~Filippenko}
\affiliation{Department of Astronomy, University of California, Berkeley, CA 94720-3411, USA}
\affiliation{Miller Senior Fellow, Miller Institute for Basic Research in Science, University of California, Berkeley, CA 94720, USA}

\author{W.~Zheng}
\affiliation{Department of Astronomy, University of California, Berkeley, CA 94720-3411, USA}

\author{S.~Bradley Cenko}
\affiliation{NASA/Goddard Space Flight Center}

\author{C.~R.~Angus}
\affiliation{Department of Physics and Astronomy, University of Southampton, Southampton, SO17 1BJ, UK}

\author{P.~J.~Brown}
\affiliation{Department of Physics and Astronomy,4242 TAMU, College Station, TX 77843, USA}
\affiliation{George P. and Cynthia Woods Mitchell Institute for Fundamental Physics \& Astronomy}

\author{T.~M.~Davis}
\affiliation{ARC Centre of Excellence for All-sky Astrophysics (CAASTRO)}
\affiliation{School of Mathematics and Physics, University of Queensland, QLD 4072, Australia}

\author{D.~A.~Finley}
\affiliation{Fermi National Accelerator Laboratory, P. O. Box 500, Batavia, IL 60510, USA}

\author{S.~Gonz\'{a}lez-Gait\'{a}n}
\affiliation{Centro de Modelamiento Matem\'{a}tico, Universidad de Chile, Beauchef 851, Piso 7, Santiago, Chile}
\affiliation{CENTRA, Instituto Superior T\'{e}cnico - Universidade de Lisboa, Portugal}

\author{C.~P.~Guti\'{e}rrez}
\affiliation{Department of Physics and Astronomy, University of Southampton, Southampton, SO17 1BJ, UK}

\author{R.~Kessler}
\affiliation{Kavli Institute for Cosmological Physics, University of Chicago, Chicago, IL 60637, USA}
\affiliation{Department of Astronomy and Astrophysics, University of Chicago, 5640 South Ellis Avenue, Chicago, IL 60637, USA}

\author{S.~Kuhlmann}
\affiliation{Argonne National Laboratory, 9700 South Cass Avenue, Lemont, IL 60439, USA}

\author{J.~Marriner}
\affiliation{Fermi National Accelerator Laboratory, P. O. Box 500, Batavia, IL 60510, USA}

\author{A.~M{\"o}ller}
\affiliation{ARC Centre of Excellence for All-sky Astrophysics (CAASTRO)}
\affiliation{The Research School of Astronomy and Astrophysics, Australian National University, ACT 2601, Australia}

\author{P.~E.~Nugent}
\affiliation{Lawrence Berkeley National Laboratory, 1 Cyclotron Road, Berkeley, CA 94720, USA}
\affiliation{Department of Astronomy, University of California, Berkeley, CA 94720-3411, USA}

\author{S.~Prajs}
\affiliation{Department of Physics and Astronomy, University of Southampton, Southampton, SO17 1BJ, UK}

\author{R.~Thomas} 
\affiliation{Lawrence Berkeley National Laboratory, 1 Cyclotron Road, Berkeley, CA 94720, USA}

\author{R. Wolf}
\affiliation{Department of Physics and Astronomy, University of Pennsylvania, Philadelphia, PA 19104, USA}

\author{A.~Zenteno}
\affiliation{Cerro Tololo Inter-American Observatory, National Optical Astronomy Observatory, Casilla 603, La Serena, Chile}

\author{T.~M.~C.~Abbott}
\affiliation{Cerro Tololo Inter-American Observatory, National Optical Astronomy Observatory, Casilla 603, La Serena, Chile}

\author{F.~B.~Abdalla}
\affiliation{Department of Physics \& Astronomy, University College London, Gower Street, London, WC1E 6BT, UK}
\affiliation{Department of Physics and Electronics, Rhodes University, PO Box 94, Grahamstown, 6140, South Africa}

\author{S.~Allam}
\affiliation{Fermi National Accelerator Laboratory, P. O. Box 500, Batavia, IL 60510, USA}

\author{J.~Annis}
\affiliation{Fermi National Accelerator Laboratory, P. O. Box 500, Batavia, IL 60510, USA}

\author{K.~Bechtol}
\affiliation{LSST, 933 North Cherry Avenue, Tucson, AZ 85721, USA}

\author{A.~Benoit-L{\'e}vy}
\affiliation{CNRS, UMR 7095, Institut d'Astrophysique de Paris, F-75014, Paris, France}
\affiliation{Department of Physics \& Astronomy, University College London, Gower Street, London, WC1E 6BT, UK}
\affiliation{Sorbonne Universit\'es, UPMC Univ Paris 06, UMR 7095, Institut d'Astrophysique de Paris, F-75014, Paris, France}

\author{E.~Bertin}
\affiliation{CNRS, UMR 7095, Institut d'Astrophysique de Paris, F-75014, Paris, France}
\affiliation{Sorbonne Universit\'es, UPMC Univ Paris 06, UMR 7095, Institut d'Astrophysique de Paris, F-75014, Paris, France}

\author{D.~Brooks}
\affiliation{Department of Physics \& Astronomy, University College London, Gower Street, London, WC1E 6BT, UK}

\author{D.~L.~Burke}
\affiliation{Kavli Institute for Particle Astrophysics \& Cosmology, P. O. Box 2450, Stanford University, Stanford, CA 94305, USA}
\affiliation{SLAC National Accelerator Laboratory, Menlo Park, CA 94025, USA}

\author{A.~Carnero~Rosell}
\affiliation{Laborat\'orio Interinstitucional de e-Astronomia - LIneA, Rua Gal. Jos\'e Cristino 77, Rio de Janeiro, RJ - 20921-400, Brazil}
\affiliation{Observat\'orio Nacional, Rua Gal. Jos\'e Cristino 77, Rio de Janeiro, RJ - 20921-400, Brazil}

\author{M.~Carrasco~Kind}
\affiliation{Department of Astronomy, University of Illinois, 1002 W. Green Street, Urbana, IL 61801, USA}
\affiliation{National Center for Supercomputing Applications, 1205 West Clark St., Urbana, IL 61801, USA}

\author{J.~Carretero}
\affiliation{Institut de F\'{\i}sica d'Altes Energies (IFAE), The Barcelona Institute of Science and Technology, Campus UAB, 08193 Bellaterra (Barcelona) Spain}

\author{F.~J.~Castander}
\affiliation{Institute of Space Sciences, IEEC-CSIC, Campus UAB, Carrer de Can Magrans, s/n,  08193 Barcelona, Spain}

\author{M.~Crocce}
\affiliation{Institute of Space Sciences, IEEC-CSIC, Campus UAB, Carrer de Can Magrans, s/n,  08193 Barcelona, Spain}

\author{C.~E.~Cunha}
\affiliation{Kavli Institute for Particle Astrophysics \& Cosmology, P. O. Box 2450, Stanford University, Stanford, CA 94305, USA}

\author{L.~N.~da Costa}
\affiliation{Laborat\'orio Interinstitucional de e-Astronomia - LIneA, Rua Gal. Jos\'e Cristino 77, Rio de Janeiro, RJ - 20921-400, Brazil}
\affiliation{Observat\'orio Nacional, Rua Gal. Jos\'e Cristino 77, Rio de Janeiro, RJ - 20921-400, Brazil}

\author{C.~Davis}
\affiliation{Kavli Institute for Particle Astrophysics \& Cosmology, P. O. Box 2450, Stanford University, Stanford, CA 94305, USA}

\author{S.~Desai}
\affiliation{Department of Physics, IIT Hyderabad, Kandi, Telangana 502285, India}

\author{H.~T.~Diehl}
\affiliation{Fermi National Accelerator Laboratory, P. O. Box 500, Batavia, IL 60510, USA}

\author{P.~Doel}
\affiliation{Department of Physics \& Astronomy, University College London, Gower Street, London, WC1E 6BT, UK}

\author{T.~F.~Eifler}
\affiliation{Department of Physics, California Institute of Technology, Pasadena, CA 91125, USA}
\affiliation{Jet Propulsion Laboratory, California Institute of Technology, 4800 Oak Grove Dr., Pasadena, CA 91109, USA}

\author{B.~Flaugher}
\affiliation{Fermi National Accelerator Laboratory, P. O. Box 500, Batavia, IL 60510, USA}

\author{P.~Fosalba}
\affiliation{Institute of Space Sciences, IEEC-CSIC, Campus UAB, Carrer de Can Magrans, s/n,  08193 Barcelona, Spain}

\author{J.~Frieman}
\affiliation{Fermi National Accelerator Laboratory, P. O. Box 500, Batavia, IL 60510, USA}
\affiliation{Kavli Institute for Cosmological Physics, University of Chicago, Chicago, IL 60637, USA}

\author{J.~Garc\'ia-Bellido}
\affiliation{Instituto de Fisica Teorica UAM/CSIC, Universidad Autonoma de Madrid, 28049 Madrid, Spain}

\author{E.~Gaztanaga}
\affiliation{Institute of Space Sciences, IEEC-CSIC, Campus UAB, Carrer de Can Magrans, s/n,  08193 Barcelona, Spain}

\author{D.~W.~Gerdes}
\affiliation{Department of Astronomy, University of Michigan, Ann Arbor, MI 48109, USA}
\affiliation{Department of Physics, University of Michigan, Ann Arbor, MI 48109, USA}

\author{D.~A.~Goldstein}
\affiliation{Department of Astronomy, University of California, Berkeley, CA 94720-3411, USA}
\affiliation{Lawrence Berkeley National Laboratory, 1 Cyclotron Road, Berkeley, CA 94720, USA}

\author{D.~Gruen}
\affiliation{Kavli Institute for Particle Astrophysics \& Cosmology, P. O. Box 2450, Stanford University, Stanford, CA 94305, USA}
\affiliation{SLAC National Accelerator Laboratory, Menlo Park, CA 94025, USA}

\author{R.~A.~Gruendl}
\affiliation{Department of Astronomy, University of Illinois, 1002 W. Green Street, Urbana, IL 61801, USA}
\affiliation{National Center for Supercomputing Applications, 1205 West Clark St., Urbana, IL 61801, USA}

\author{J.~Gschwend}
\affiliation{Laborat\'orio Interinstitucional de e-Astronomia - LIneA, Rua Gal. Jos\'e Cristino 77, Rio de Janeiro, RJ - 20921-400, Brazil}
\affiliation{Observat\'orio Nacional, Rua Gal. Jos\'e Cristino 77, Rio de Janeiro, RJ - 20921-400, Brazil}

\author{G.~Gutierrez}
\affiliation{Fermi National Accelerator Laboratory, P. O. Box 500, Batavia, IL 60510, USA}

\author{K.~Honscheid}
\affiliation{Center for Cosmology and Astro-Particle Physics, The Ohio State University, Columbus, OH 43210, USA}
\affiliation{Department of Physics, The Ohio State University, Columbus, OH 43210, USA}

\author{D.~J.~James}
\affiliation{Astronomy Department, University of Washington, Box 351580, Seattle, WA 98195, USA}

\author{M.~W.~G.~Johnson}
\affiliation{National Center for Supercomputing Applications, 1205 West Clark St., Urbana, IL 61801, USA}

\author{K.~Kuehn}
\affiliation{Australian Astronomical Observatory, North Ryde, NSW 2113, Australia}

\author{N.~Kuropatkin}
\affiliation{Fermi National Accelerator Laboratory, P. O. Box 500, Batavia, IL 60510, USA}

\author{T.~S.~Li}
\affiliation{Fermi National Accelerator Laboratory, P. O. Box 500, Batavia, IL 60510, USA}

\author{M.~Lima}
\affiliation{Departamento de F\'isica Matem\'atica, Instituto de F\'isica, Universidade de S\~ao Paulo, CP 66318, S\~ao Paulo, SP, 05314-970, Brazil}
\affiliation{Laborat\'orio Interinstitucional de e-Astronomia - LIneA, Rua Gal. Jos\'e Cristino 77, Rio de Janeiro, RJ - 20921-400, Brazil}

\author{M.~A.~G.~Maia}
\affiliation{Laborat\'orio Interinstitucional de e-Astronomia - LIneA, Rua Gal. Jos\'e Cristino 77, Rio de Janeiro, RJ - 20921-400, Brazil}
\affiliation{Observat\'orio Nacional, Rua Gal. Jos\'e Cristino 77, Rio de Janeiro, RJ - 20921-400, Brazil}

\author{J.~L.~Marshall}
\affiliation{Department of Physics and Astronomy,4242 TAMU, College Station, TX 77843, USA}
\affiliation{George P. and Cynthia Woods Mitchell Institute for Fundamental Physics \& Astronomy}

\author{P.~Martini}
\affiliation{Center for Cosmology and Astro-Particle Physics, The Ohio State University, Columbus, OH 43210, USA}
\affiliation{Department of Astronomy, The Ohio State University, Columbus, OH 43210, USA}

\author{F.~Menanteau}
\affiliation{Department of Astronomy, University of Illinois, 1002 W. Green Street, Urbana, IL 61801, USA}
\affiliation{National Center for Supercomputing Applications, 1205 West Clark St., Urbana, IL 61801, USA}

\author{C.~J.~Miller}
\affiliation{Department of Astronomy, University of Michigan, Ann Arbor, MI 48109, USA}
\affiliation{Department of Physics, University of Michigan, Ann Arbor, MI 48109, USA}

\author{R.~Miquel}
\affiliation{Instituci\'o Catalana de Recerca i Estudis Avan\c{c}ats, E-08010 Barcelona, Spain}
\affiliation{Institut de F\'{\i}sica d'Altes Energies (IFAE), The Barcelona Institute of Science and Technology, Campus UAB, 08193 Bellaterra (Barcelona) Spain}

\author{R.~L.~C.~Ogando}
\affiliation{Laborat\'orio Interinstitucional de e-Astronomia - LIneA, Rua Gal. Jos\'e Cristino 77, Rio de Janeiro, RJ - 20921-400, Brazil}
\affiliation{Observat\'orio Nacional, Rua Gal. Jos\'e Cristino 77, Rio de Janeiro, RJ - 20921-400, Brazil}

\author{D.~Petravick}
\affiliation{National Center for Supercomputing Applications, 1205 West Clark St., Urbana, IL 61801, USA}

\author{A.~A.~Plazas}
\affiliation{Jet Propulsion Laboratory, California Institute of Technology, 4800 Oak Grove Dr., Pasadena, CA 91109, USA}

\author{A.~K.~Romer}
\affiliation{Department of Physics and Astronomy, Pevensey Building, University of Sussex, Brighton, BN1 9QH, UK}

\author{E.~S.~Rykoff}
\affiliation{Kavli Institute for Particle Astrophysics \& Cosmology, P. O. Box 2450, Stanford University, Stanford, CA 94305, USA}
\affiliation{SLAC National Accelerator Laboratory, Menlo Park, CA 94025, USA}

\author{M.~Sako}
\affiliation{Department of Physics and Astronomy, University of Pennsylvania, Philadelphia, PA 19104, USA}

\author{E.~Sanchez}
\affiliation{Centro de Investigaciones Energ\'eticas, Medioambientales y Tecnol\'ogicas (CIEMAT), Madrid, Spain}

\author{V.~Scarpine}
\affiliation{Fermi National Accelerator Laboratory, P. O. Box 500, Batavia, IL 60510, USA}

\author{R.~Schindler}
\affiliation{SLAC National Accelerator Laboratory, Menlo Park, CA 94025, USA}

\author{M.~Schubnell}
\affiliation{Department of Physics, University of Michigan, Ann Arbor, MI 48109, USA}

\author{I.~Sevilla-Noarbe}
\affiliation{Centro de Investigaciones Energ\'eticas, Medioambientales y Tecnol\'ogicas (CIEMAT), Madrid, Spain}

\author{R.~C.~Smith}
\affiliation{Cerro Tololo Inter-American Observatory, National Optical Astronomy Observatory, Casilla 603, La Serena, Chile}

\author{M.~Soares-Santos}
\affiliation{Fermi National Accelerator Laboratory, P. O. Box 500, Batavia, IL 60510, USA}

\author{F.~Sobreira}
\affiliation{Instituto de F\'isica Gleb Wataghin, Universidade Estadual de Campinas, 13083-859, Campinas, SP, Brazil}
\affiliation{Laborat\'orio Interinstitucional de e-Astronomia - LIneA, Rua Gal. Jos\'e Cristino 77, Rio de Janeiro, RJ - 20921-400, Brazil}

\author{E.~Suchyta}
\affiliation{Computer Science and Mathematics Division, Oak Ridge National Laboratory, Oak Ridge, TN 37831}

\author{M.~E.~C.~Swanson}
\affiliation{National Center for Supercomputing Applications, 1205 West Clark St., Urbana, IL 61801, USA}

\author{G.~Tarle}
\affiliation{Department of Physics, University of Michigan, Ann Arbor, MI 48109, USA}

\author{A.~R.~Walker}
\affiliation{Cerro Tololo Inter-American Observatory, National Optical Astronomy Observatory, Casilla 603, La Serena, Chile}

\collaboration{(The DES Collaboration)}
\noaffiliation
\email{mat.smith@soton.ac.uk}

\begin{abstract}
We present observations of \nm, the first spectroscopically confirmed hydrogen-free superluminous supernova (SLSN-I) at redshift $z\approx2$. \nm\ was discovered by the Dark Energy Survey (DES) Supernova Program, with follow-up photometric data from the \textit{Hubble Space Telescope}, Gemini, and the European Southern Observatory Very Large Telescope supplementing the DES data. Spectroscopic observations confirm \nm\ to be at $z=1.998$, and spectroscopically similar to Gaia16apd (a SLSN-I at $z=0.102$), with a peak absolute magnitude of $U=-22.26\pm0.06$. The high redshift of \nm\ provides a unique opportunity to study the ultraviolet (UV) properties of SLSNe-I. Combining \nm\ with ten similar events from the literature, we show that there exists a homogeneous class of SLSNe-I in the UV ($\lambda_\mathrm{rest}\approx2500$\,\AA), with peak luminosities in the (rest-frame) $U$ band, and increasing absorption to shorter wavelengths. There is no evidence that the mean photometric and spectroscopic properties of SLSNe-I differ between low ($z<1$) and high redshift ($z>1$), but there is clear evidence of diversity in the spectrum at $\lambda_\mathrm{rest}<2000$\,\AA, possibly caused by the variations in temperature between events. No significant correlations are observed between spectral line velocities and photometric luminosity. Using these data, we estimate that SLSNe-I can be discovered to $z=3.8$ by DES. While SLSNe-I are typically identified from their blue observed colors at low redshift ($z<1$), we highlight that at $z>2$ these events appear optically red, peaking in the observer-frame $z$-band. Such characteristics are critical to identify these objects with future facilities such as the Large Synoptic Survey Telescope, \textit{Euclid}, and the \textit{Wide-Field Infrared Survey Telescope}, which should detect such SLSNe-I to $z=3.5$, 3.7, and 6.6, respectively.
\end{abstract}


\section{Introduction}
\label{sec:Intro}

Superluminous supernovae (SLSNe) are extremely bright, but rare, supernova-like events, peaking at luminosities in excess of $M_\mathrm{AB}\approx-21$\,mag, around a hundred times greater than classical core-collapse events. While at least two classes of SLSNe have been observationally identified \citep[see][]{GalYam2012,Quimby2011, Inserra2017}, the most common are known as SLSNe-I, or SLSNe-Ic, as they are hydrogen poor \citep{Quimby2011,Inserra2013} and spectroscopically similar to Type Ic supernovae. The power source of SLSNe-I remains unclear, but it is unlikely to be the radioactive decay of \nickel\ that powers the light curves of normal SNe \citep[see][]{Chatzopoulos2009,Chomiuk2011,Inserra2013,Papadopoulos2015}. Alternative mechanisms could be the spin down of a rapidly-rotating  magnetar \citep[e.g.,][]{Kasen2010,Woosley2010,Bersten2016}, a pair-instability SN explosion \citep[e.g.,][]{Langer2007,Kasen2011,Kozyreva2017}, the interaction of SN ejecta with circumstellar material \citep[e.g.,][]{Woosley2007,Chevalier2011,Chatzopoulos2012,Chatzopoulos2013}, or the fallback of accreted material onto a compact remnant \citep{Dexter2013}.

Beyond the puzzle of their astrophysical nature, SLSNe are ideal probes of the high-redshift universe because of their high luminosity and long-lived (time-dilated) light curves. With the peak of their energy output in the ultraviolet (UV), redshifted into the optical at high redshift, studies \citep[e.g.,][]{Cooke2012,Berger2012,Howell2013,Pan2017} have detected a few SLSN candidates at $z>1.5$, and suggest that the SLSN rate at such redshifts is a factor of $\sim15$ higher than at low redshifts, possibly tracking the cosmic star-formation rate \citep{Prajs2017}. Therefore, there is significant potential for detecting many more SLSNe to $z\approx 4$ and beyond with forthcoming facilities like the Large Synoptic Survey Telescope \citep[LSST;][]{Scovacricchi2016}, \textit{Euclid} \citep{Inserra2017b}, and the \textit{Wide-Field Infrared Survey Telescope} \citep[\textit{WFIRST};][]{Yan2017}. 

High-redshift observations of SLSNe-I can provide additional constraints on the physical processes that drive these events. A typical spectrum of a SLSN-I around maximum light can be well described by a hot, black-body continuum  with broad absorption lines produced by C, O, Si, Mg, and Fe with characteristic velocities of $10{,}000$ to $20{,}000$ \kms\ \citep{Quimby2011,Inserra2013}. These lines are especially prominent in the UV spectra of SLSNe-I, with a characteristic turnover at $\sim2900$\,\AA\ where strong line-blanketing from iron-group elements suppresses the black-body flux. These features provide information on the synthesized material and intrinsic metal abundance of the progenitor star, necessary to constrain the physical processes that govern SLSNe-I.  

\citet{Yan2017}, using a \textit{Hubble Space Telescope} (\textit{HST}) spectrum of Gaia16apd at $z=0.102$, showed that SLSNe-I exhibit significantly less line blanketing in the UV than any classical-luminosity SN, and hence are likely metal-poor in the outer ejecta. However, given the scarsity of low-redshift SLSNe-I and difficulties of observing in the observer-frame UV, Gaia16apd and the recently discovered SN~2017egm \citep{Dong2017_SN2017egm,Bose_egm2017} are the only SLSNe-I in the local Universe ($z<0.3$) with rest-frame UV spectral information. At $z>1$, observations at optical wavelengths probe the rest-frame UV, presenting an avenue to determine the diversity of SLSNe-I in the UV, and thus provide constraints on progenitor scenarios. 

In this paper we present \nm, a new SLSN-I discovered by the Dark Energy Survey (DES) Supernova Program \citep[DES-SN;][]{Bernstein2012}. At $z=1.998$, \nm\ is the highest-redshift spectroscopically confirmed SN (of any type) to date. The two SLSN candidates of \cite{Cooke2012} found in Canada-France-Hawaii Telescope Legacy Survey data were not spectroscopically confirmed in real time, but their redshifts ($z=2.05$ and 3.90) were measured from host-galaxy spectra. \nm\ also supersedes the highest-redshift spectroscopically confirmed Type Ia supernovae (SNe~Ia). For example, \citet{rodney2012} provide strong evidence for a spectroscopically confirmed SN Ia at $z=1.55$ (``SN Primo") detected as part of the \textit{HST} CANDELS project. Classifications of higher-redshift SNe~Ia rely on other information, such as model fits to the photometric light-curve data and/or prior knowledge of the host-galaxy redshift; thus, they do not possess independent spectroscopic confirmations \citep[e.g.,][]{riess2001,jones2013,rodney2014,rodney2016,Rubin2017}.

In Section~\ref{sec:observations}, we provide the observational data used to detect and classify \nm, while in Section~\ref{sec:analysis} we analyze the UV properties of this SLSN, and compare it to the literature. Section~\ref{sec:discussion} discusses the consequences of this discovery for ongoing and future searches of SLSNe. We summarize our conclusions in Section~\ref{sec:conclusions}. Throughout, we assume H$_0=70$\,km\,s$^{-1}$\,Mpc$^{-1}$ and a flat $\Lambda$CDM cosmology with $\Omega_\mathrm{matter}=0.3$.

\section{Observations}
\label{sec:observations}

\nm\ was discovered at $\alpha=03^{\mathrm{h}}40^{\mathrm{m}}14\fs83$ and $\delta=-29\arcdeg05\arcmin53\farcs 5$ (J2000) in the first DECam \citep{FlaugherDECam2015} images of the fourth season (Y4) of DES-SN \citep{Bernstein2012}. It was detected with the difference imaging pipeline \citep{Kessler2015} in all passbands ($griz$) taken on 2016 August 22 (all dates herein are UTC) and was not detected in the last DES-SN images taken in the previous season (Y3) on 2016 February 08. \nm\ did, however, have a marginal detection at $3.3\sigma$ (point source) in $g$-band DECam images taken as part of SUDSS\footnote{``Search Using DECam for Superluminous Supernovae'' (PI: Sullivan)} on 2016 March 9, but was not detected in any other SUDSS image taken at that time. \nm\ was first designated a transient on 2016 August 27 and monitored in all passbands ($griz$) on an average 7 day cadence by DES-SN until the end of January 2017, when it fell below the detection limit of the single-epoch DES-SN images. Further details of the DES-SN observing strategy and spectroscopic follow-up programs can be found in \citet{DAndreaSpectroscopy}. Information on the DES-SN difference-imaging search pipeline and machine-learning algorithms to identify transient objects can be found in \citet{Kessler2015} and \citet{Goldstein2015}.

\begin{figure}
\centering
\includegraphics[width=0.95\textwidth]{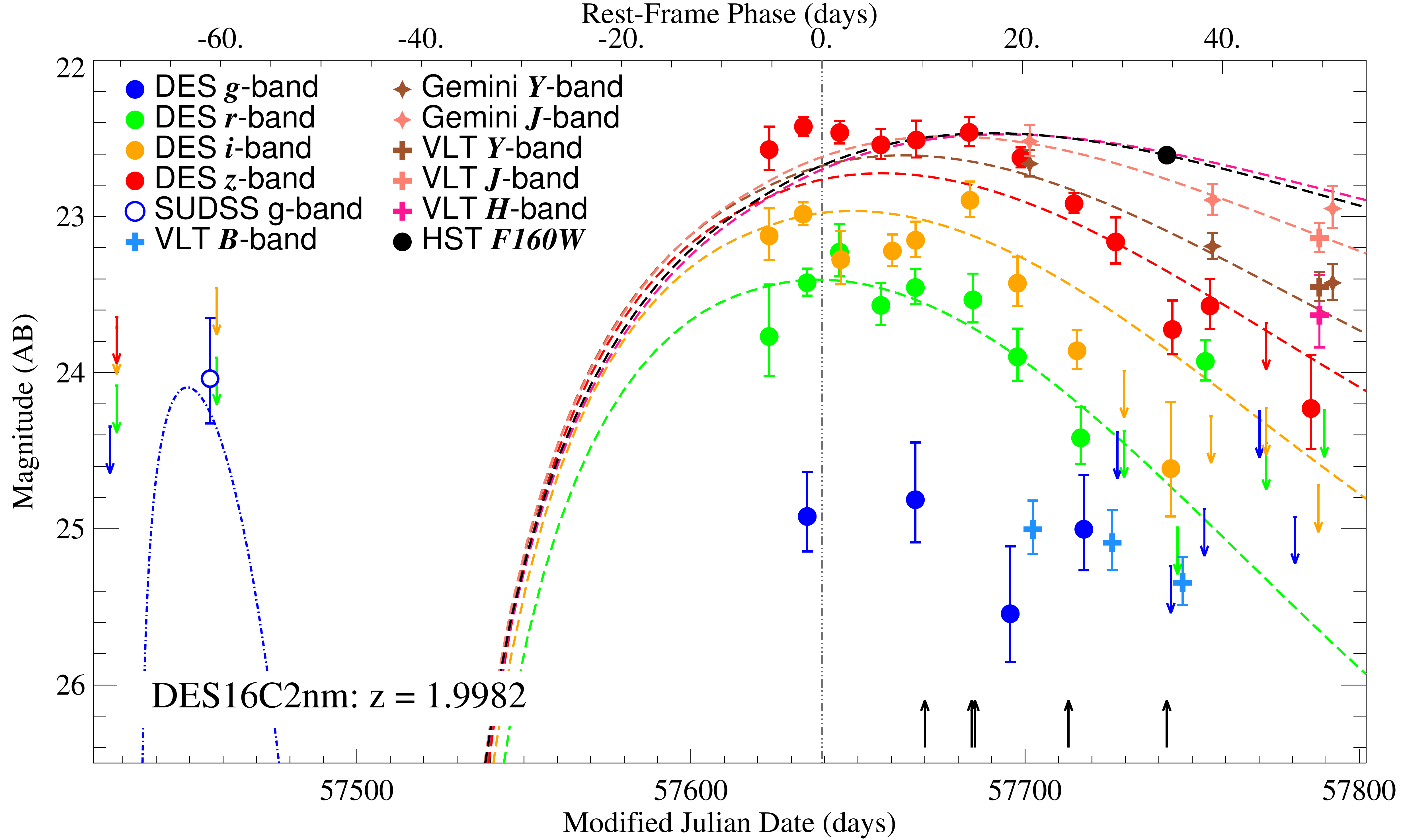}
\caption{The photometric data of \nm\ from DES (filled circles), SUDSS $g$ band (open circle), Gemini/Flamingoes-2 (stars), \textit{HST} (black filled circle), VLT/FORS2 and VLT/HAWKI (crosses). The DES data have been averaged with a 14-day window for clarity.  3$\sigma$ upper limits are shown as downward arrows, and the various epochs of spectroscopy are shown as black upward arrows. The best-fit magnetar model evaluated at $z=1.9982$ (Section~\ref{sec:physical-modelling}) is shown as dashed lines, with the model of \citet{Piro2015} at early times (Section~\ref{sec:early-time}) plotted as dash-dotted lines. The colors correspond to different passbands. The phases presented are relative to peak luminosity in the observer-frame DES $i$ band (MJD = 57639) as highlighted by a black dot-dashed line.
\label{fig:c2nm_lc}}
\end{figure}

Photometric measurements were made using the pipeline discussed by \citet{Papadopoulos2015} and \citet{Smith2016}, which has also been extensively used in the literature \citep[e.g.,][and references therein]{PTFphot}. This pipeline subtracts a deep template image from each individual SN image to remove the host-galaxy light using a point-spread-function (PSF) matching routine. SN photometry is then measured from the difference image using a PSF fitting technique. Figure~\ref{fig:c2nm_lc} shows the full light curve of \nm\ averaged with a 14-day window for clarity.

The time of maximum light, which varies with wavelength, can be best constrained from our $i$-band data. A third order polynomial fit to the observer-frame fluxes indicates that \nm\ reached a peak magnitude in the $i$ band of $23.12\pm0.08$ \citep[after correcting for foreground extinction using $E(B-V)=0.014$\,mag;][]{Schlegel1998} on MJD 57639$\pm$14. \nm\ peaked 33$\pm$21 days earlier at $23.43\pm0.10$ in the $r$ band, and 9$\pm$7 days later at $22.48\pm0.04$ in the $z$ band. The uncertainties on the time of maximum light are driven by the slow photometric evolution of \nm. In the $g$ band, most of our detections have low signal-to-noise ratio (S/N). Based on the available data, we estimate the $g$-band peak magnitude to be prior to the start of DES Y4 (MJD = 57623.4). The individual photometric measurements are listed in Table~\ref{tab:lc_table} and are also available from the WISeREP archive\footnote{\url{http://wiserep.weizmann.ac.il/}} \citep{WISEREP}. 

\nm\ was prioritized for follow-up spectroscopy based on its slowly evolving light curve (in both brightness and color). The first spectrum was obtained on 2016 October 09 \citep[][when the object was $i\approx23.0$\,mag]{C2nm_ATEL} at the Magellan-Clay Telescope at the Las Campanas Observatory, using the Low Dispersion Survey Spectrograph (LDSS-3). The spectrum was reduced using standard \texttt{IRAF} routines\footnote{Image Reduction and Analysis Facility,  distributed by the National Optical Astronomy Observatory, which is operated by the Association of Universities for Research in Astronomy (AURA) under a cooperative agreement with the U.S. National Science Foundation.}, and has an effective wavelength 
coverage of 4500 to $10{,}000$\,\AA\ (observer frame). This spectrum showed evidence of the \ion{Mg}{2} absorption doublet (2796\,\AA\ and 2804\,\AA\ rest frame) from the host galaxy at 8387\,\AA\ and 8403\,\AA, but at low S/N.

\begin{figure}
\centering
\includegraphics[width=0.95\textwidth]{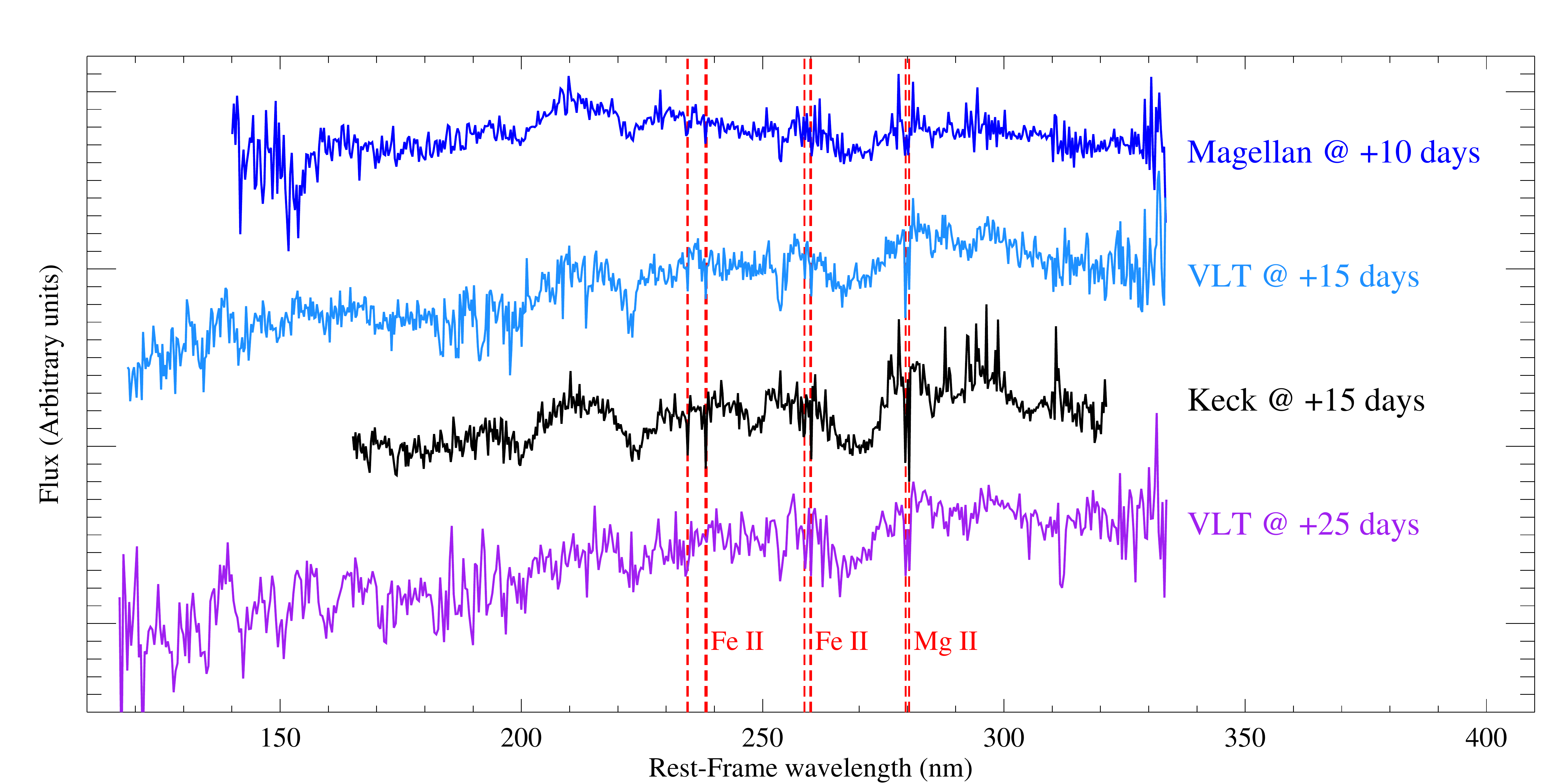}
\caption{The spectral series of \nm, with rest-frame phase information (relative to peak luminosity in the observer-frame $i$ band (MJD = 57639)). The spectra have been smoothed (to 10\,\AA, 8\,\AA, 8\,\AA, and 12\,\AA, respectively), adjusted to match the colors from the nearest epoch of $griz$ photometry, and are offset for clarity. Prominent host-galaxy absorption features are labeled in red. 
\label{fig:c2nm_spectra}}
\end{figure}

\nm\ was then spectroscopically reobserved at the European Southern Observatory (ESO) Very Large Telescope (VLT) using the X-SHOOTER instrument on 2016 October 24 and 25, as well as the Keck-II telescope using the DEep Imaging Multi-Object Spectrograph \citep[DEIMOS;][]{Faber2003} instrument on 2016 October 25. These spectra unambiguously confirmed the presence of the \ion{Mg}{2} doublet at 8382\,\AA\ and 8402\,\AA, as well as additional absorption features produced by \ion{Fe}{2} (2344, 2383, 2587, 2600\,\AA\ rest frame). These features confirmed the redshift of \nm\ as $z=1.9982\pm0.0004$, where the uncertainty is the observed scatter in the redshift estimates obtained from the individual absorption features. Figure~\ref{fig:c2nm_spectra} shows the observer-frame optical spectral series of \nm\ including an additional spectrum taken at the VLT on 2016 November 22, once the high redshift of \nm\ was known. 

\begin{figure*}
\centering
\includegraphics[width=0.95\textwidth]{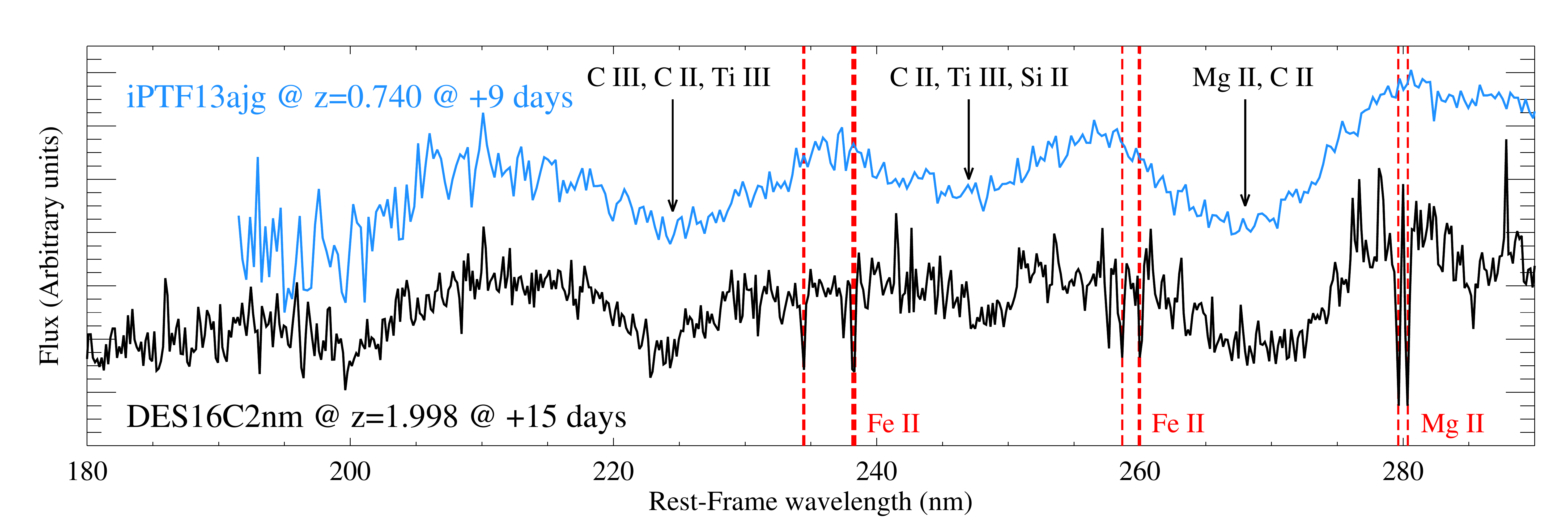}
\caption{The rest-frame optical spectrum of \nm\ obtained at Keck-II on 2016 October 25. Prominent galaxy absorption features due to \ion{Mg}{2} and \ion{Fe}{2} are labeled in red. Also shown is a spectrum of iPTF13ajg \citep[$z=0.740$;][]{Vreeswijk2014} at a similar phase in the light curve. A velocity shift of 3{,}000\,km\,s$^{-1}$ has been applied to the iPTF13ajg spectrum as discussed in Section~\ref{subsec:uvspec}. Key spectral features from the SN identified by \citet{Mazzali2016} and discussed in Section~\ref{subsec:uvspec} are highlighted with downward arrows. 
\label{fig:c2nm_spectra_redshift}}
\end{figure*}

Figure~\ref{fig:c2nm_spectra_redshift} shows the Keck spectrum of \nm\ 15 rest-frame days after maximum light in $i$-band, highlighting the galaxy absorption features used to determine the redshift. For comparison, we also show the spectrum of iPTF13ajg \citep[a SLSN-I at $z=0.740$][]{Vreeswijk2014} at approximately the same phase (9 days after maximum light) in the light curve. The broad absorption
features at 2200\,\AA, 2450\,\AA, and 2700\,\AA\ (rest frame) are common in SLSN-I spectra around maximum light \citep{Quimby2011,Vreeswijk2014,Mazzali2016} and spectroscopically classify \nm\ as a SLSN-I at $z=1.9982$. 

To constrain the rest-frame optical properties of \nm, a near-infrared (NIR) spectrum using WFC3/G141 on HST was obtained on 2016 December 20. Information about all spectra can be found in Table~\ref{tab:spec_table}. The spectra are also available from WISeREP.

Once the high redshift of \nm\ was established, further follow-up observations, including NIR imaging from \textit{HST}, Gemini/Flamingos-2, and VLT/HAWK-I and optical imaging from VLT/FORS2 (detailed in Table~\ref{tab:lc_ancilliary_table}), were obtained. Reductions of the \textit{HST}, VLT/HAWK-I, and VLT/FORS2 data were performed using the standard pipelines provided by the various observatories, with the Gemini/Flamingos-2 data reduced using the THELI pipeline \citep{Theli_Erben2005,Theli_Schirmer2013}, following standard procedures outlined by \citet{Theli_Schirmer2013}. Magnitudes were estimated using \texttt{SEXTRACTOR} \citep{Bertin1996}. These ancillary photometric data are shown in Figure~\ref{fig:c2nm_lc}. 

The host galaxy of \nm\ is detected in stacked images from DES Science Verification (SV) that contain no SN light. Using \texttt{SEXTRACTOR}, we measure host-galaxy AB magnitudes (\texttt{MAG\_AUTO}) of $g$, $r$, $i$, $z$ = $24.98\pm0.13$, $25.64\pm0.32$, $24.85\pm0.19$, and $24.42\pm0.17$, respectively (after correcting for Galactic extinction). Using the \texttt{Z-PEG} photometric-redshift software \citep{LeBorgne2002}, we estimate the stellar mass of the host galaxy to be $\log(M/\msolar)=10.1^{+0.7}_{-1.2}$ and a star-formation rate (SFR) of  $\log(\mathrm{SFR}/\msolar\,\mathrm{yr}^{-1})=1.3^{+0.2}_{-0.9}$, assuming a \citet{Kroupa2001} initial mass function and fixing the redshift at $z=1.998$. The host galaxy of \nm\ is not detected in any existing NIR catalogues, necessary to reliably constrain the continuum flux and metallicity at $z=2$. Despite this, these host-galaxy properties are consistent with the properties of other, lower-redshift SLSN-I host galaxies as seen in DES \citep{Papadopoulos2015,Smith2016,Pan2017} and in the literature \citep[e.g.,][]{Neill2011,Lunnan2014,Leloudas2015,Perley2016,Angus2016,Chen2017}.

\section{Analysis}
\label{sec:analysis}

The high redshift of \nm\ provides an opportunity to study the rest-frame UV properties of SLSNe-I, and, by comparing to literature events, to study their diversity and possible evolution with redshift. We first consider the photometric properties of \nm\ compared with other SLSNe-I in the literature. Phases given are relative to peak luminosity in the observer-frame $i$ band (MJD = 57639).

\subsection{Rest-Frame UV Light Curve} 
\label{sec:rest-frame}

\begin{figure}
\centering
\includegraphics[width=0.95\textwidth]{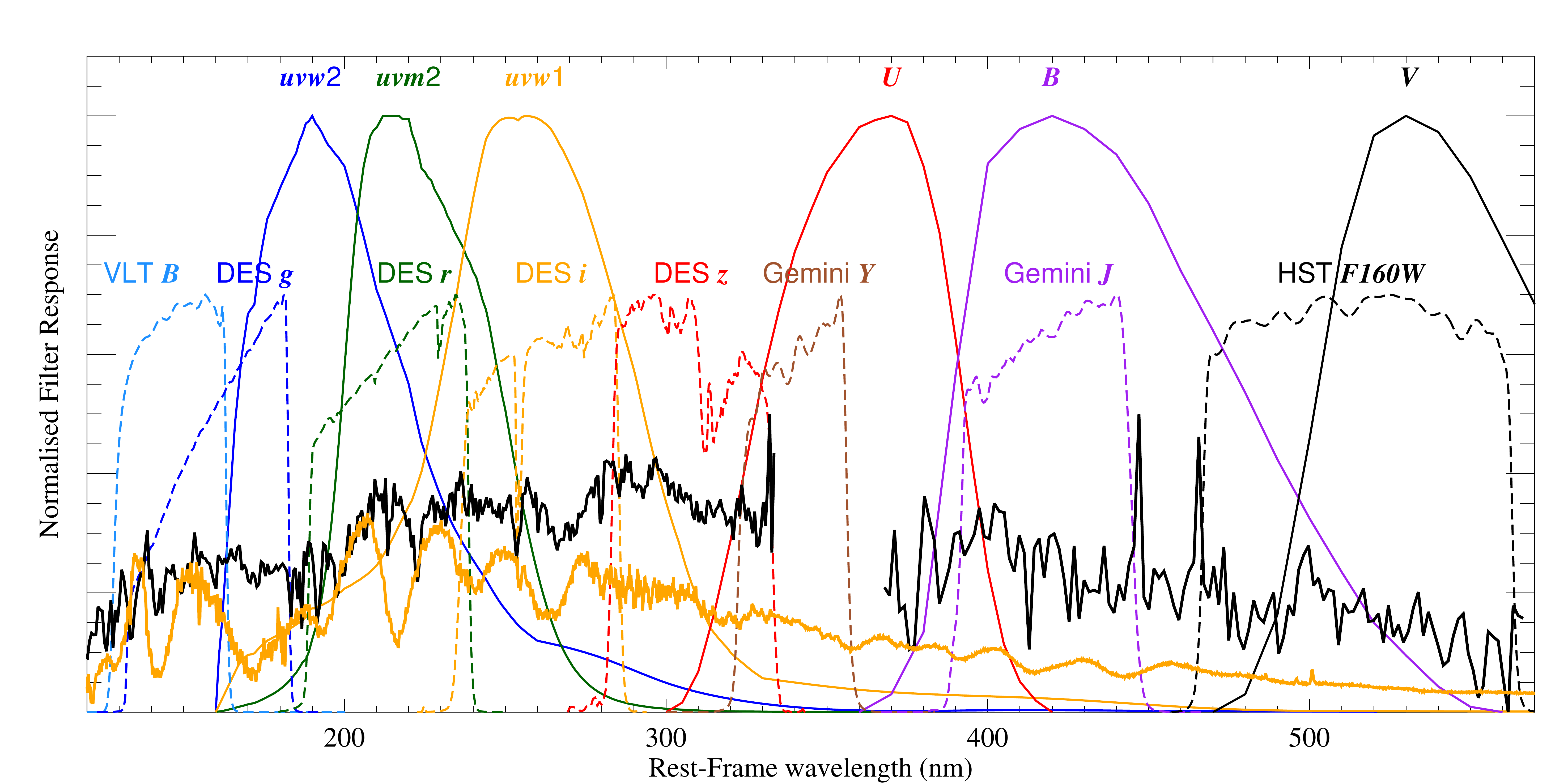}
\caption{The rest-frame wavelength coverage for our observations of \nm. Our observed $g$, $r$, $i$, $z$, $Y$, $J$, and $F160W$ filters are shown as dashed lines but shifted back to the rest frame assuming $z=1.9982$. The solid lines are the rest-frame  \textit{Swift} $uvw2$, $uvm2$, $uvw1$, and Johnson $U$, $B$, $V$ filters. Our VLT and \textit{HST} spectra of \nm\ (at +15\,d and +34\,d, respectively) are shown in black with the composite spectrum of Gaia16apd (at maximum light) shown in orange. The filter pairs used for the cross-filter $K$-corrections (e.g., the DES $i$ band $\rightarrow$ $uvw1$) are plotted in the same color. The DES $z$ band is used to correct to rest-frame $U$ owing to the larger dataset around maximum light. 
\label{fig:c2nm_filter_response}}
\end{figure}

We use a standard $K$-correction procedure to estimate the rest-frame light curve of \nm. Figure~\ref{fig:c2nm_filter_response} shows the rest-frame wavelength range covered by our data collected on \nm, compared to the rest-frame ($z=0$) \textit{Swift} \citep{Gehrels2004, Roming2005, Poole2008, Breeveld2011} and Johnson \citep{Bessell1990} filters. This figure illustrates that the rest-frame $uvw2$ ($\lambda_{\mathrm{eff}}=2080$\,\AA), $uvm2$ (2255\,\AA), $uvw1$ (2615\,\AA), $U$ (3605\,\AA), $B$ (4413\,\AA), and $V$ (5512\,\AA) passbands roughly correspond to the observer-frame $g$, $r$, $i$, $z$, $J$, and $H$ passbands (respectively) for \nm. As our VLT and \textit{HST} spectra have limited wavelength coverage and do not overlap, we use the spectra of Gaia16apd \citep{Yan2017} at maximum light for $K$-corrections to the light curve of \nm. This spectrum is similar to that of \nm\ and is the only available spectrum of a SLSN-I that covers the full spectral range probed by the photometry of \nm\ (1200 to 8700\,\AA; Figure~\ref{fig:c2nm_filter_response}). To test the validity of this approach, we measure the scatter in the inferred absolute magnitude of \nm\ when using the maximum-light spectrum of Gaia16apd compared to the spectra of \nm, PS1-11bam \citep{Berger2012}, iPTF13ajg \citep{Vreeswijk2014}, SNLS06-D4eu \citep{Howell2013}, SCP06F6 \citep{Barbary2009}, and Gaia16apd \citep{Yan2017} after maximum light. These are the only literature SLSNe-I with spectra extending below 2000\,\AA in the rest-frame. We find standard deviations of 0.39 (in the $uvw2$ filter), 0.08 ($uvm2$), 0.16 ($uvw1$), 0.15 ($U$), 0.01 ($B$), and 0.14 ($V$)\,mag between the $K$-corrections determined using the spectrum of Gaia16apd at peak and the available literature spectra. No evidence of increased scatter is seen as a function of phase. These $K$-correction uncertainties are added in quadrature to the measured uncertainties of the rest-frame light curve of \nm\ and comparison literature events. Early-phase data ($<60$ days pre-peak) are $K$-corrected using a black body of $27{,}000$\,K, as discussed in Section~\ref{sec:early-time}.

\begin{figure}
\centering
\includegraphics[width=0.95\textwidth]{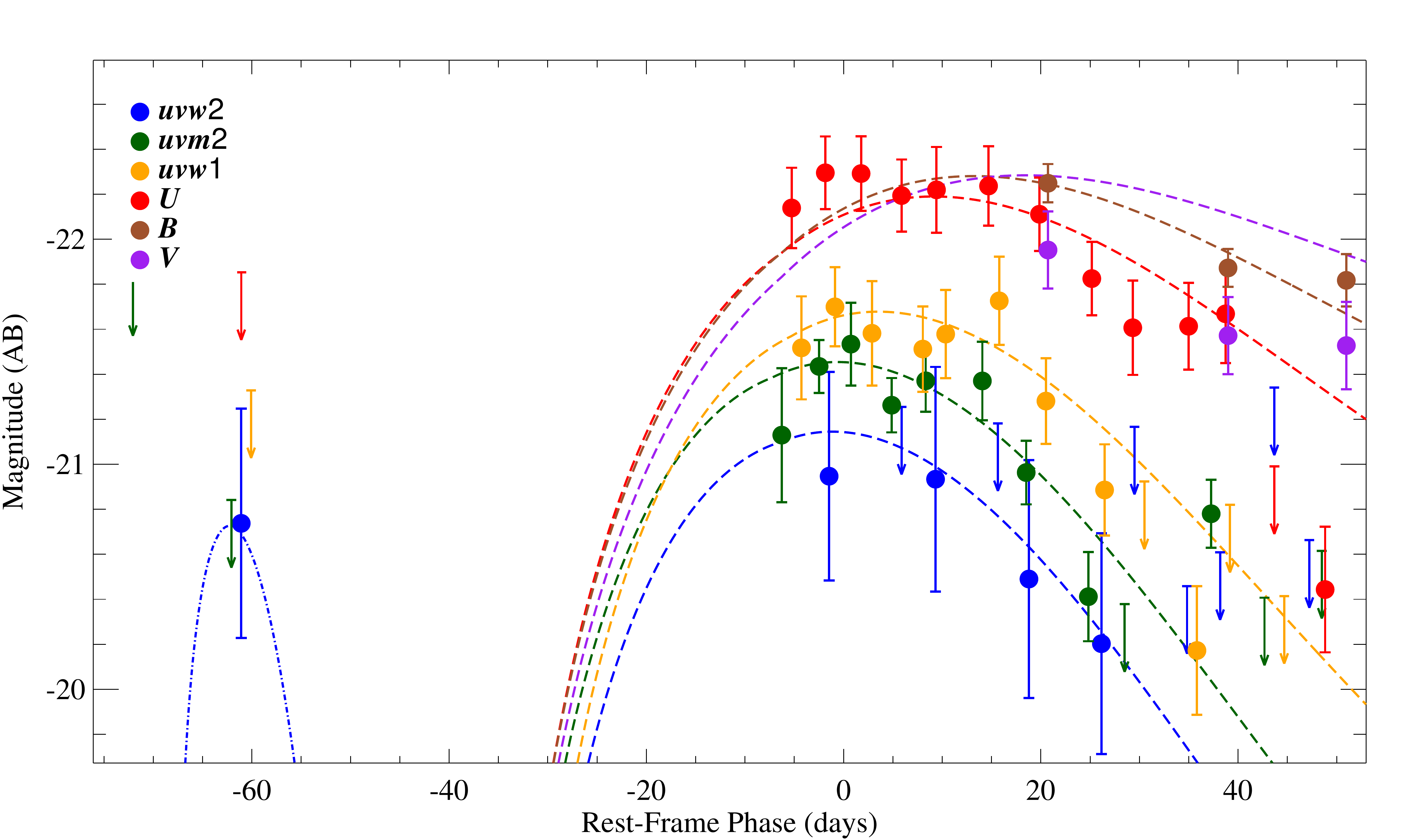}
\caption{The rest-frame ($uvw2$, $uvm2$, $uvw1$, $U$, $B$, $V$) light curves of \nm. Magnitudes are estimated with a standard $K$-correction procedure, using the spectrum of Gaia16apd for the main light curve and a black body of 27{,}000\,K for the early-time data (see Sections~\ref{sec:rest-frame} and~\ref{sec:early-time}). Uncertainties are determined from the observer-frame light curve and include a filter-dependent $K$-correction uncertainty (see Section~\ref{sec:rest-frame}) added in quadrature. 3$\sigma$ upper limits are denoted with arrows. For clarity, the $uvm2$ and $uvw1$ data points have been offset by $-1$ and +1 day, respectively. The best-fit magnetar model (Section~\ref{sec:physical-modelling}) is shown as dashed lines, with the model of \citet{Piro2015} at early times (Section~\ref{sec:early-time}) plotted as a dash-dotted line. 
\label{fig:c2nm_restframe}}
\end{figure}

At each epoch in the main light curve, we adjust the spectrum of Gaia16apd so that its synthetic photometry matches the observed photometry of \nm\ at $z=1.9982$ and calculate the cross-filter $K$-correction required to determine the rest-frame magnitudes in the filters listed above. We fit a polynomial function to the resulting rest-frame light curves, shown in Figure~\ref{fig:c2nm_restframe}, and determine peak absolute magnitudes of $uvm2=-21.41\pm0.10$, $uvw1=-21.69\pm0.07$, and $U=-22.28\pm0.04$ (AB). \nm\ is most luminous (in integrated flux) in the rest-frame $U$ band, and is one of the brightest SLSNe-I (at peak) discovered so far; only iPTF13ajg \citep{Vreeswijk2014} is brighter at peak ($U=-22.47$\,mag) using the same methodology. The phase of maximum light is statistically consistent for all three filters. We do not have sufficiently high-quality data around maximum light in the rest-frame $B$ and $V$ bands to make similar measurements. 
 
\subsection{Physical Modeling} 
\label{sec:physical-modelling}

The light curves of SLSNe-I are well described by the spin-down of a rapidly rotating magnetar \citep{Inserra2013,Papadopoulos2015,Smith2016}. We fit this model \citep[as described in Appendix D of][]{Inserra2013} to the observer-frame fluxes of \nm\ around maximum light ($-20<$ phase $<60$ days), fitting all filters simultaneously and show the best-fitting magnetar model in Figures~\ref{fig:c2nm_lc} and~\ref{fig:c2nm_restframe}. The best-fit model has an initial spin period of $P_\mathrm{ms}=1.33\pm0.18$\,ms, an initial magnetic field strength of $B_{14}=0.68\pm0.07\times 10^{14}$\,G, and a diffusion timescale of $\tau_\mathrm{m}=39.62\pm8.28$\,days assuming an opacity of $\kappa=0.1$ and an explosion energy of $10^{51}$erg. From these parameter values, we infer an ejecta mass of $M_\mathrm{ej}=6.23\pm2.45\,\msolar$. The best fitting model has $\chi_{\textrm{red}}^2=4.4$, suggesting that while the model is a good match to the light-curve and color evolution of \nm\, it does not capture the full variation, potentially due to the strong UV absorption observed in the spectral series of \nm\  (Figure~\ref{fig:c2nm_spectra}). These inferred parameters are consistent with the modeling of other SLSNe-I in the literature, such as SN2011kf \citep[e.g.][]{Nicholl2017}. 

\subsection{Early-Time Data} 
\label{sec:early-time}

\nm\ was first observed with a significance of 3.3$\sigma$ in SUDSS $g$-band imaging and 2.7$\sigma$ in $r$-band imaging on 2016 March 9, 61 rest-frame days prior to maximum light.  No other 3$\sigma$ detections are found in the previous 2\,yr of DES imaging.  While this detection is at low significance,  it is consistent in phase with the precursor peaks observed in other SLSNe \citep{Smith2016,Nicholl2015,Leloudas2012}. In particular, LSQ14bdq \citep{Nicholl2015} showed a bright and relatively fast precursor peak, 60 rest-frame days prior to the peak of the main SLSN. These precursors have peak temperatures of $\sim25{,}000$\,K and cool rapidly, with typical timescales of 15 rest-frame days \citep{Smith2016}, and are well modeled as shock cooling from extended material around the progenitor star \citep{Ofek2010,Piro2015}. 

Considering a black body, the observer-frame $g$ and $r$-band fluxes of the precursor detection are best fit by a temperature and radius of $25{,}000$\,K and $7\times{10}^{14}$\,cm, respectively, consistent with values determined from other literature events \citep{Smith2016,Nicholl2015} and when all available photometric measurements ($g$, $r$, $i$, and $z$ band) are considered. Finally, in Figure~\ref{fig:c2nm_restframe}, we compare the rest-frame luminosity of this epoch with a shock-cooling model from \citet{Piro2015} with parameters consistent with those obtained for the light curve of LSQ14bdq \citep{Smith2016}. For this epoch, $K$-corrections are determined using the best-fitting black body with $27{,}000$\,K. The precursor epoch  has a peak luminosity comparable to the main light curve in $uvw2$ ($\lambda_{\textrm{eff}}=2600$\,\AA) and is consistent with the \citet{Piro2015} model of shock cooling from extended material (dash-dotted line). 

\subsection{Comparison to Literature Events}
\label{sec:fuv_luminosity}

\begin{figure}
\centering
\includegraphics[width=0.95\textwidth]{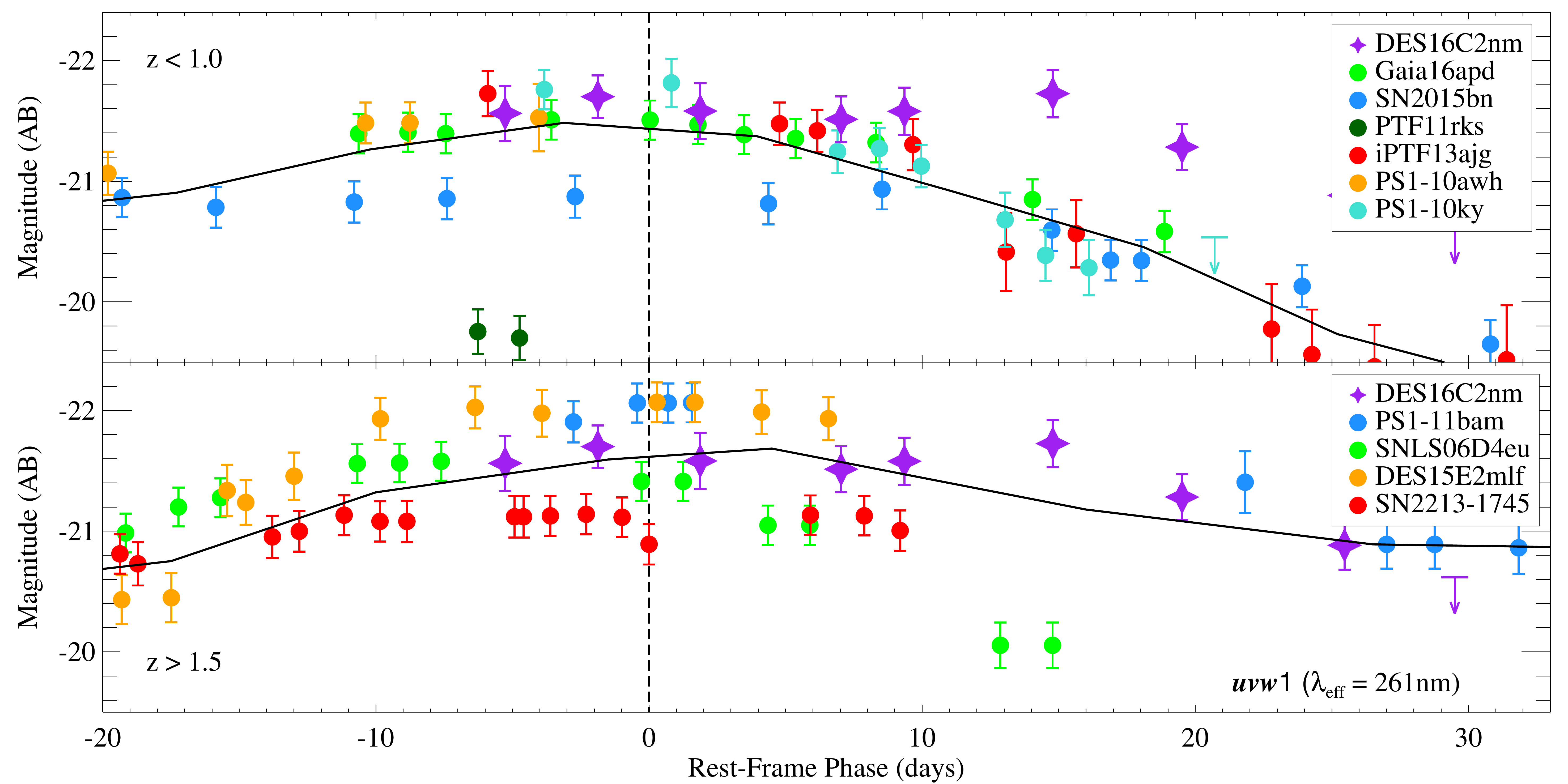}
\caption{The $uvw1$ light curve of DES16C2nm compared to literature events at low ($z<1$; top panel) and high ($z>1$; lower panel) redshifts. $K$-corrections have been determined using the spectrum of Gaia16apd around maximum light as discussed in the text. An average of the available light curves is overplotted in black. Data for the literature objects are from various sources \citep{Chomiuk2011,Berger2012,Cooke2012,Inserra2013,Howell2013,Vreeswijk2014,Nicholl2016a,Yan2017,Pan2017}.
\label{fig:c2nm_restframe_comp}}
\end{figure}

\begin{figure}
\centering
\includegraphics[width=0.95\textwidth]{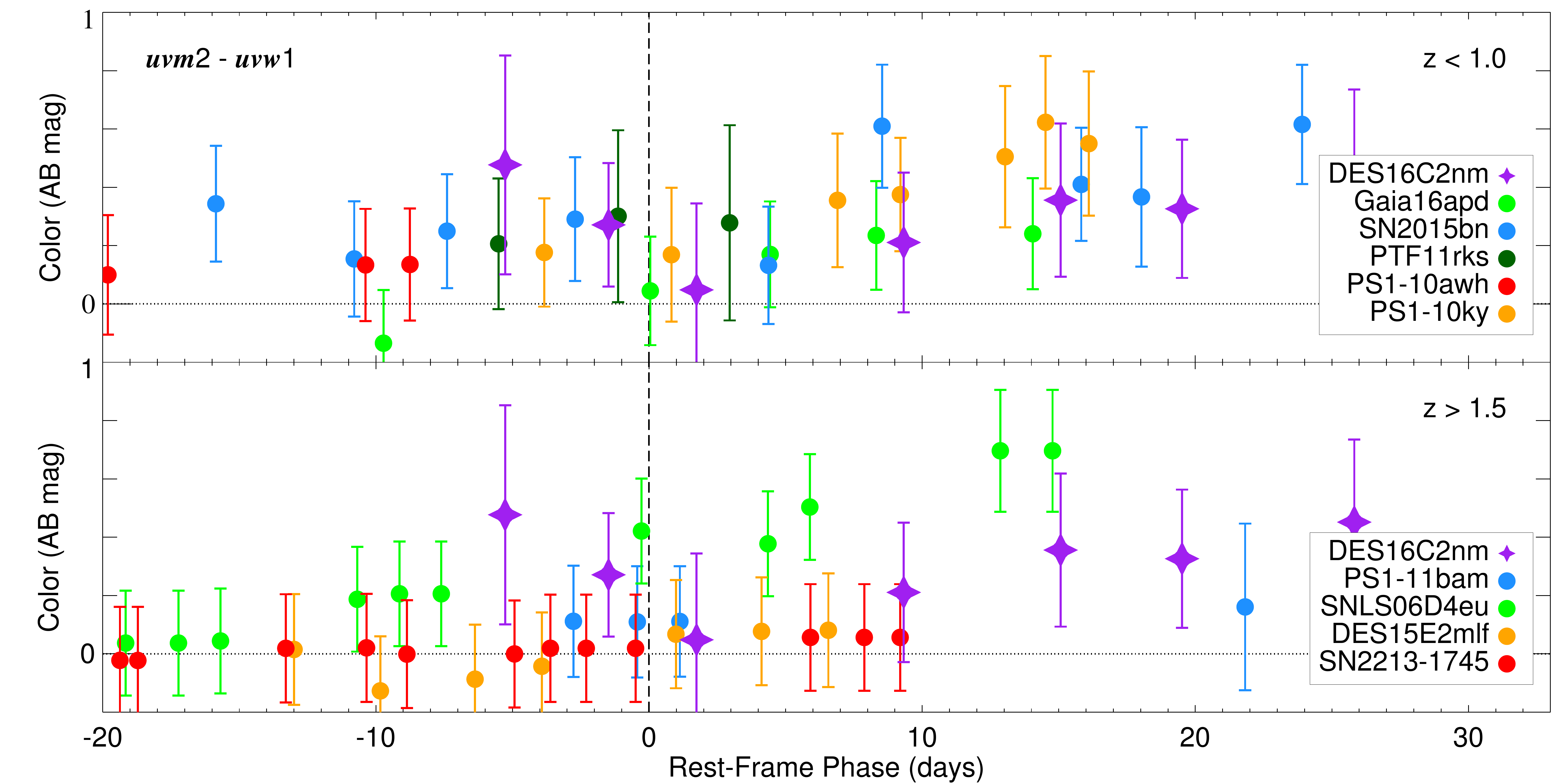}
\caption{The rest-frame $uvm2-uvw1$ color evolution of DES16C2nm compared to literature events at low ($z<1$; top panel) and high ($z>1$; lower panel) redshifts. The color is determined by requiring observations in two distinct filters, as discussed in the text. Data for the literature objects are from various sources \citep{Chomiuk2011,Berger2012,Cooke2012,Inserra2013,Howell2013,Nicholl2016a,Yan2017,Pan2017}.
\label{fig:c2nm_restframe_comp_color}}
\end{figure}

In Figures~\ref{fig:c2nm_restframe_comp} and~\ref{fig:c2nm_restframe_comp_color}, we compare the rest-frame UV light curves of \nm\ to SLSNe-I in the literature. Figure~\ref{fig:c2nm_restframe_comp} compares the rest-frame \textit{Swift} $uvw1$ light curve of \nm\ to SLSNe-I at low ($z<1$) and high ($z>1$) redshift. All SLSNe-I have been $K$-corrected with the procedure described for \nm\ using the observed filter having an effective wavelength closest to 2615\,\AA\ (the central wavelength of \textit{Swift} $uvw1$). Uncertainties due to our $K$-correction procedure are included for all events.

While the number of low-redshift SLSNe-I is small owing to the low intrinsic rate of these events and the limited volume probed, there is evidence of a homogeneous sample of SLSNe-I with similar peak luminosity distributions and evolution. The UV luminosity (at $\lambda_{\mathrm{eff}}=2615$\,\AA) of \nm\ is consistent with other SLSNe-I in the literature, including Gaia16apd at $z=0.102$. The exceptions are PTF11rks, which is known to be underluminous in all filters \citep{Inserra2013}, and SN~2015bn, which has been shown to evolve more slowly (photometrically and spectroscopically) than other events \citep{Nicholl2016a}. The remaining objects exhibit a scatter of $<0.3$\,mag at peak. 

There is some evidence that the light curve of \nm\ is broader than other literature events, but this may be caused by uncertainties in the estimate of maximum light (4 rest-frame days; Section~\ref{sec:observations}), as the light curve shows little evolution around maximum light at these wavelengths. Further, with observations preferentially in the rest-frame optical, the time of maximum light for low-redshift SLSNe-I is typically estimated in the rest-frame $U$ band. \nm\ peaks 14$\pm$7 days after our fiducial estimate of maximum light in $U$-band.  At high redshift ($z>1$), the scatter in the luminosity distribution of the SLSNe-I is larger, but all events have $-22.1 < M_{uvw1} < -21.1$\,mag at peak (i.e., a scatter of $<1$\,mag). This includes the photometrically classified SN2213-1745 \citep{Cooke2012}, which has a host-galaxy spectroscopic redshift of 2.05. No evidence of a difference in the mean peak luminosity of SLSNe-I at high and low redshifts is observed, with $M_\textrm{peak}=-21.61\pm0.48$ and $M_\textrm{peak}=-21.40\pm0.29$\,mag, respectively. 

Figure~\ref{fig:c2nm_restframe_comp_color} shows the rest-frame \textit{Swift} $uvm2-uvw1$ color evolution of SLSNe-I at low ($z<1$) and high ($z>1$) redshifts. In this analysis, 
we follow the procedure described above, obtaining $K$-corrections for $uvm2$ and $uvw1$ from different observed filters. In all cases, SLSNe-I show little scatter in their rest-frame UV color near maximum light, with a scatter of $<1$\,mag, and they exhibit little evidence of significant color evolution ($\Delta \text{color} < 1$\,mag) around maximum light ($\pm10$ days). 
At all redshifts, the majority of SLSNe-I have $uvm2-uvw1>0$\,mag at maximum light. From this analysis, there is no indication that SLSNe-I at high redshift exhibit different colors, or evolve differently, from those at lower redshifts. 

\subsection{Rest-Frame UV Spectral Evolution} 
\label{subsec:uvspec}

\begin{figure}
\centering
\includegraphics[width=0.90\textwidth]{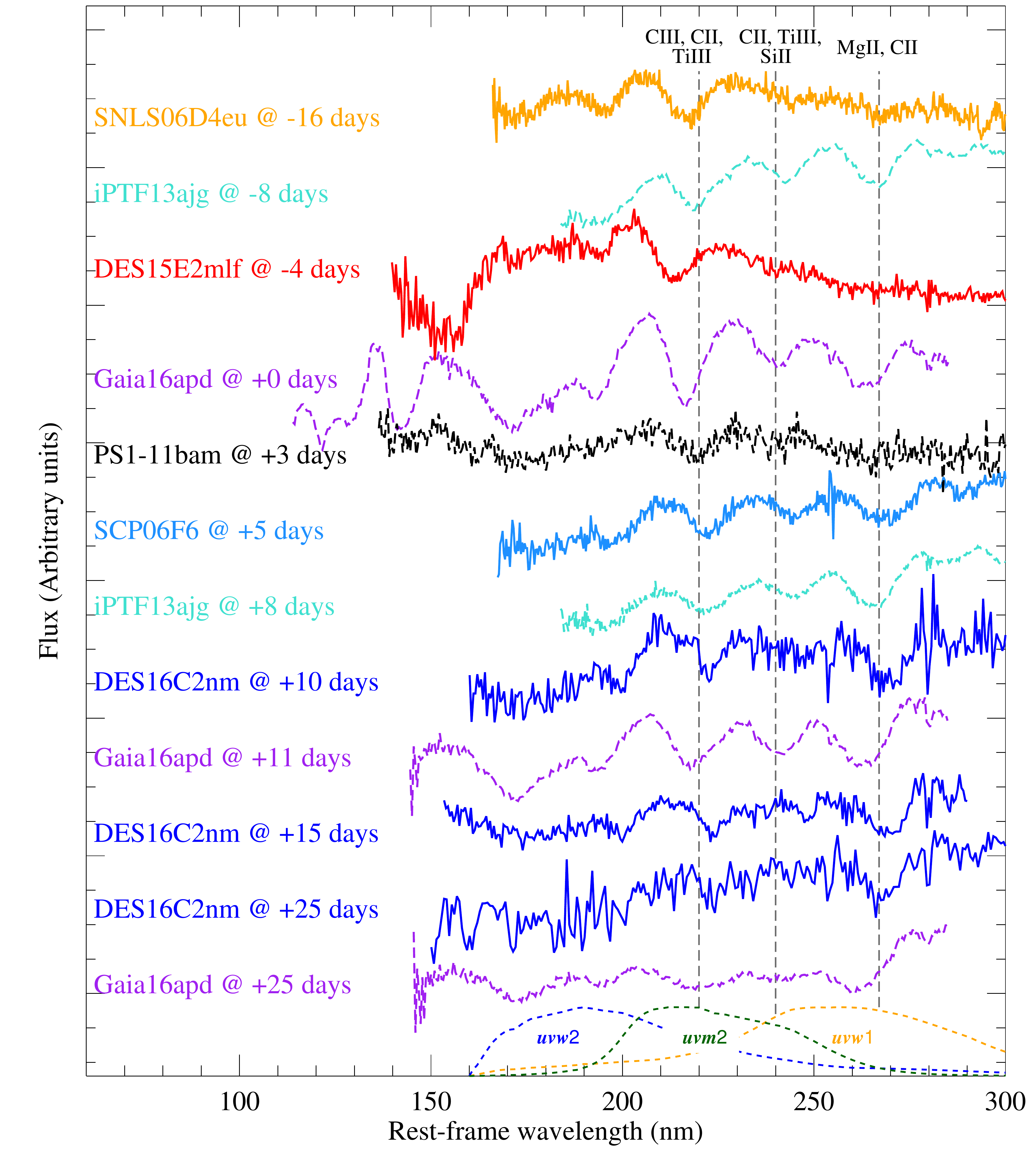}
\caption{The FUV spectroscopic evolution of SLSNe-I. A rest-frame UV spectral time series of \nm\ and compiled literature SLSNe-I as a function of phase from maximum light. Low-redshift ($z<1$) SLSNe-I are plotted with dashes, while high-redshift ($z>1$) events are shown as solid lines. The spectra have been smoothed and offset for clarity. Key spectral features from the SNe identified by \citet{Mazzali2016} and discussed in Section~\ref{subsec:uvspec} at 267, 240, and 220\,nm are highlighted with vertical lines. \textit{Swift} filter responses are also shown to highlight the effective wavelength coverage probed by these spectra compared to the photometric analysis in Figures~\ref{fig:c2nm_restframe_comp} and ~\ref{fig:c2nm_restframe_comp_color}. Data for the literature objects are from various sources \citep{Barbary2009,Berger2012,Howell2013,Vreeswijk2014,Yan2017,Pan2017}.
\label{fig:c2nm_spectra_comp}}
\end{figure}

We now consider the spectral diversity and evolution of SLSNe-I compared to \nm. Figure~\ref{fig:c2nm_spectra_comp} shows a temporal series of UV spectra of \nm\ compared to other UV spectral data from the literature. As optical observations at $z>1$ probe the rest-frame UV spectrum, the vast majority of SLSNe-I with available rest-frame UV data around maximum light are at $z>1$, and in particular $z>1.5$. At $z<1$, only Gaia16apd has rest-frame spectral data below 1800\,\AA. 

At $\lambda>2000$\,\AA, the vast majority of SLSNe-I exhibit strong absorption features at $\sim2200$\,\AA\ \citep[identified as a blend of \ion{C}{3}, \ion{C}{2}, and \ion{Ti}{3} by][]{Mazzali2016}, at $\sim2400$\,\AA\ (identified as \ion{C}{2}, \ion{Ti}{3}, and \ion{Si}{2}), and at $\sim2650$\,\AA\ (identified as \ion{Mg}{2} and \ion{C}{2}). These features are ubiquitous to all SLSNe-I, with the exception of SNLS-06D4eu and DES15E2mlf, which only exhibit absorption at $\sim2200$\,\AA. These spectra are modeled by \citet{Mazzali2016} to have higher temperatures ($\sim18{,}000$\,K) than canonical events (such as iPTF13ajg, with $T\approx12{,}000$\,K). At these temperatures, in the \citet{Mazzali2016} model, \ion{C}{2} and \ion{Mg}{2} are ionised to \ion{C}{3} and \ion{Mg}{3}, respectively. The absence of \ion{C}{2} and \ion{Mg}{2} reduces absorption at $\sim2400$\,\AA\ and $\sim2650$\,\AA, but the presence of \ion{C}{3} preserves the absorption feature at $\sim2200$\,\AA. The higher intrinsic temperature of these SLSNe results in higher expansion velocities, resulting in the observed absorption features being blueshifted relative to other events. With higher intrinsic temperatures, the black bodies for these events peak at bluer wavelengths than for lower-temperature SLSNe, such as iPTF13ajg.

For events where absorption at $\sim2200$\,\AA, $\sim2400$\,\AA, and $\sim2650$\,\AA\ is seen, there is no evidence of evolution or significant diversity in the strength of these lines around maximum light, consistent with the $uvw1$ luminosity distribution determined in  Figure~\ref{fig:c2nm_restframe_comp}. The spectra of \nm\ are consistent with those of literature SLSNe-I, with some evidence of a lower strength in the $\sim2400$\,\AA\ absorption feature. However, Figure~\ref{fig:c2nm_spectra_comp} suggests diversity in the observed velocity distribution of SLSNe-I. \cite{Liu2017} measured the photospheric velocity evolution of SLSNe-I, from the \ion{Fe}{2} 5169\,\AA\ feature. They show that SLSNe velocities range from $10{,}000$ to $20{,}000$ \kms\ around maximum light, before falling linearly at phases greater than 10 days after peak. This  result is consistent with that found by \citet{Chomiuk2011} and \citet{Inserra2013}, and matches the dispersion observed in Figure~\ref{fig:c2nm_spectra_comp}. At comparable phases, \nm\ shows a lower velocity than Gaia16apd and reveals no evidence of velocity evolution in the $\sim2650$\,\AA\ feature (see also Figure~\ref{fig:c2nm_spectra}), but is consistent with the velocity evolution of other events, such as iPTF13ajg. Considering all available spectra, we find no evidence of a correlation between line velocity and far-UV (FUV) luminosity or color (see Section~\ref{sec:fuv_luminosity}). 

Below $\lambda<2000$\,\AA\ all events exhibit absorption at $\sim1950$\,\AA, determined to be a blend of \ion{Fe}{3} and \ion{Co}{3}, and an additional feature at $\sim1700$\,\AA, tentatively identified as \ion{Al}{3} and \ion{Si}{3} by \citet{Mazzali2016}.  For SLSNe-I with higher intrinsic temperatures, such as SNLS-06D4eu and DES15E2mlf, the black body peaks in this wavelength range. There is no evidence of varying absorption at $\lambda<$2000\,\AA\, so the diversity in this wavelength range is primarily driven by differences in the underlying temperature distribution. 

\section{Discussion}
\label{sec:discussion}

\subsection{Rest-Frame UV: Implications for High-Redshift Searches} 

The discovery and spectroscopic confirmation of SLSNe-I at $z=2$ provides unique insights into how to identify such events at higher redshifts. With lower metallicities and younger stellar populations at $z>2$, these studies are key to distinguishing whether SLSNe-I show diversity with respect to the metallicity, mass, or age of the stellar population, thereby constraining progenitor scenarios. Observations of SLSNe-I at $z>2$ open an avenue to measuring the interstellar medium \citep[ISM;][]{Berger2012} in distant galaxies, measuring cosmological parameters beyond the reach of SNe~Ia \citep{Inserra2014}, and finding the explosions of Population III stars \citep{Cooke2012}. 

At low redshift, SLSNe-I are typically identified by their blue color, driven by their spectral energy distribution (SED) peaking in the rest-frame $U$ band, and by their long-duration light curves. However, as discussed in Section~\ref{subsec:uvspec}, at higher redshifts, the $U$-band flux is redshifted toward redder filters, and the rest-frame UV and FUV become prominent in the selection. At $z=2$, the rest-frame $U$-band flux is no longer visible in optical filters (Figure~\ref{fig:c2nm_filter_response}), and only the UV flux is observed in optical searches. Figure~\ref{fig:c2nm_restframe} characterizes the UV properties of \nm, highlighting that in the UV, significant absorption and a turnover in the black body result in reduced flux in the FUV, compared to the optical, with decreasing flux toward the Lyman limit. The consequence of this is that at high redshift ($z>1.5$), SLSNe-I have a characteristic red observer-frame color and long-duration light curves. 

\begin{figure}
\centering
\includegraphics[width=0.90\textwidth]{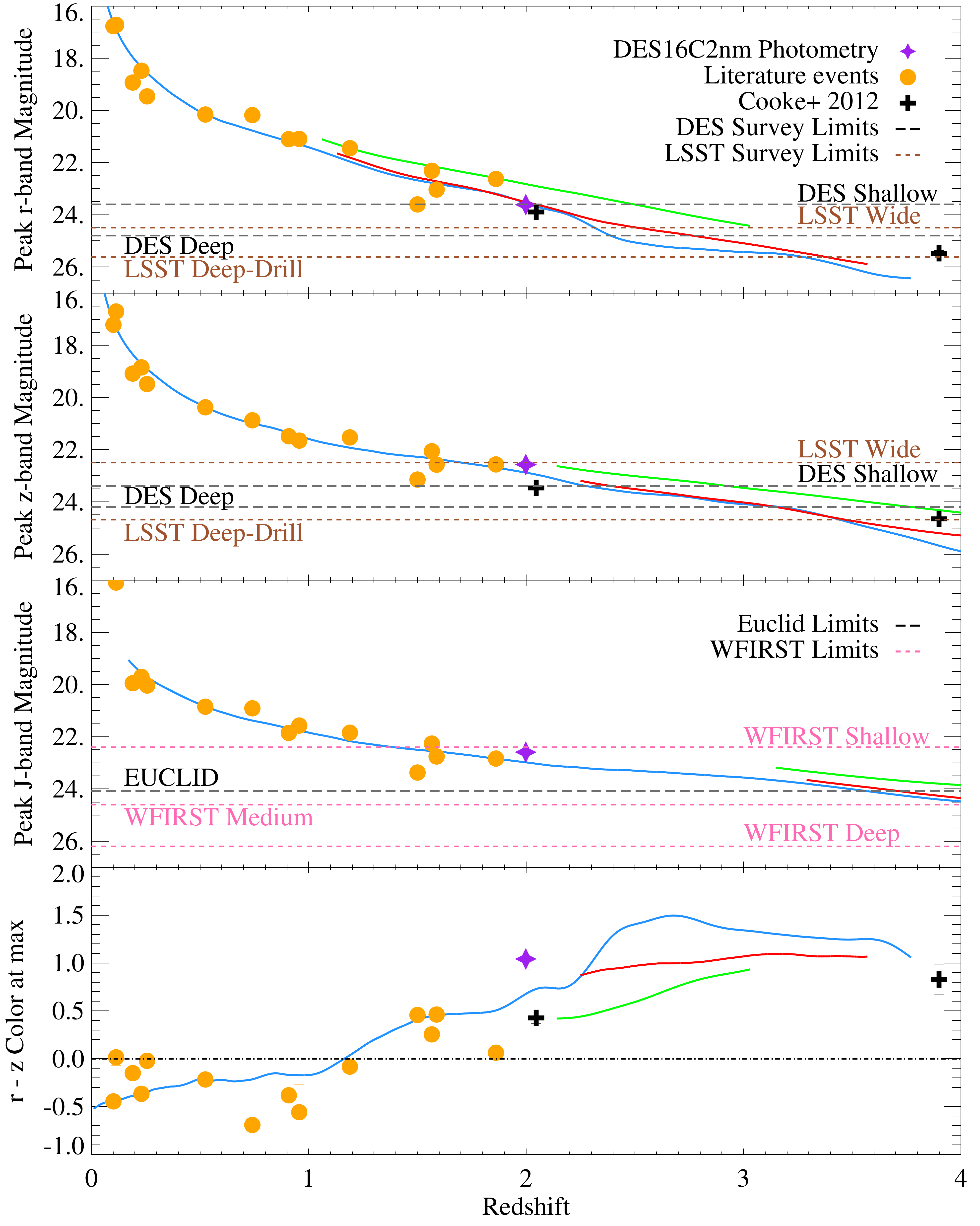}
\caption{The observer-frame evolution of SLSNe-I. \textit{Upper panel}: The apparent $r$-band evolution of SLSNe-I 
as a function of redshift. \nm\ is shown as a purple star, with literature SLSNe-I in orange and the photometrically 
confirmed objects of \citet{Cooke2012} shown as black crosses. Values determined from the maximum-light 
spectrum of Gaia16apd, normalized to the peak magnitude of Gaia16apd, are shown in blue, with the 
corresponding \nm\ spectrum from VLT in red and DES15E2mlf in green. $K$-corrections for literature events 
have been determined using the Gaia16apd spectrum. The magnitude limits for the DES deep and shallow 
fields and LSST wide-field and deep-drilling fields are also shown. \textit{Second panel}: As above, only for the $z$ band. \textit{Third panel}: The $J$-band evolution of SLSNe-I as a function of redshift. The magnitude limits for the \textit{Euclid} and \textit{WFIRST} shallow, medium, and deep surveys are also shown. \textit{Lower panel}: The $r-z$ color evolution 
of SLSNe-I as a function of redshift as inferred from the upper two panels.  A horizontal line of  $r-z=0$\,mag highlights the difference between low ($z<1$) and high ($z>1$) redshift events. 
\label{fig:slsn_mag_col_evol}}
\end{figure}

To identify SLSNe-I at $z>2$ requires information on the UV spectrum and evolution. As highlighted in Section~\ref{subsec:uvspec}, data in this wavelength range are sparse, with only \nm, Gaia16apd, PS1-11bam, and DES15E2mlf having data that extend to $\lambda<1700$\,\AA. Figure~\ref{fig:slsn_mag_col_evol} shows the luminosity and color evolution of SLSNe-I at maximum light inferred from these spectra as a function of redshift. The spectra have been normalized and adjusted to the peak magnitude and color of each event. These spectra are then redshifted between 0 and 4, and integrated through the DECam $r$ and $z$ passbands. Figure~\ref{fig:slsn_mag_col_evol} shows the evolution in luminosity and color for these spectra, along with other events in the literature. Where required, $K$-corrections have been determined from the spectrum of Gaia16apd. The single-epoch 5$\sigma$ magnitude limits of the DES deep and shallow fields are shown along with those expected for LSST wide-field and deep-drilling fields \citep{Scovacricchi2016,ScolnicPrivate}. These limits suggest that in a single epoch, DES can detect SLSNe-I that are spectroscopically similar to Gaia16apd and \nm, to $z=3.2$ in $z$-band observations of its deep fields or $z=3.8$ for objects that are spectroscopically similar to DES15E2mlf. For the LSST deep-drilling fields, the maximum redshifts in the $z$ band are 3.5 for the \nm\ spectrum and 4.3 when considering the DES15E2mlf spectrum. 

This analysis can be extended to the observer-frame NIR, where surveys such as \textit{Euclid} \citep[with 5$\sigma$ limiting magnitude per visit $J\approx24.1$ in the deep fields and an irregular cadence of 10 visits to each  of two fields in a 6-month observing window for 5\,yr;][]{Inserra2017b} and \textit{WFIRST} \citep[limiting magnitude per exposure, $J=22.4$, 24.6,  and 26.2, and a regular cadence of 5 days;][]{Hounsell2017} will search for high-redshift SLSNe-I. An object spectrally similar to \nm\ can be detected to $z=3.7$ with \textit{Euclid} and $z=1.4$, 4.4, and 6.6 in the \textit{WFIRST} shallow, medium, and deep surveys, respectively. 

From the lower panel of Figure~\ref{fig:slsn_mag_col_evol}, SLSNe-I in the literature are usually characterized as being observationally blue at low redshift, with all but one event having $r-z<0$\,mag at $z<1$. This is well matched by the optical and NIR spectrum of Gaia16apd. At intermediate redshifts ($z\approx1.5$), where the $z$ band probes the peak of the black-body emission in the rest-frame $U$ band, and the $r$ band probes the absorption triplet at $\lambda\approx2500$\,\AA, SLSNe-I in the literature are red with $0.0<r-z<0.5$\,mag. The sole exception is the intrinsically hot, and therefore bluer, DES15E2mlf, which still has $r-z>0$\,mag. 
 
The lower panel of Figure~\ref{fig:slsn_mag_col_evol} provides a diagnostic to identify SLSNe-I at $z>2$. In this redshift range, the available spectra, which are warped to be consistent with available photometry, all indicate that SLSNe-I have $r-z>0.5$\,mag at maximum light, with increasing color as a function of redshift. This observation is in good agreement with \nm\ and the photometrically classified SN2213-1745 at $z=2.05$ \citep{Cooke2012}. 

At $z>2$, SLSNe-I will have characteristic observer-frame timescales of $\sim6$ months, with little luminosity evolution over this time. This, combined with the magnitude and color evolution described above, will enable the robust identification of these events from existing datasets, such as DES, HSC, PS1, SNLS, and SDSS, and upcoming facilities such as LSST, \textit{Euclid}, and \textit{WFIRST}. Given the slow evolution of SLSNe-I and time-dilation effects at high redshift, these estimates are conservative. Current and upcoming surveys will be able to combine single-epoch images over month-long timescales to obtain deep stacks with which to identify SLSNe-I to $z>4$. 

\section{Conclusions}
\label{sec:conclusions}

The first confirmation of a SLSN-I at $z=2$ provides new insights into these ultraluminous events. The spectroscopic similarity between \nm\ at $z=1.9982$, iPTF13ajg at $z=0.740$, and Gaia16apd at $z=0.102$ highlights that these events exist at all redshifts, and show little evolution with cosmic time. Comparing the UV luminosity and color evolution of SLSNe-I at a range of redshifts, we find no evidence of a difference in mean luminosity or color between SLSNe-I at high and low redshift. After maximum light, these events show similar spectroscopic features in the UV in the range 2000--3000\,\AA.

The enormous luminosity of these events means that it is theoretically possible to discover and monitor them to $z>4$. However, at $z>1.5$ identifying these objects requires an understanding of the FUV spectra of SLSNe-I, for which only a handful of objects exist. In its first 4\,yr of operations, DES has discovered and spectroscopically confirmed the two highest-redshift SLSNe-I to date. These objects show differing levels of absorption in the rest-frame UV at $\lambda<2000$\,\AA, but the implications from both events are clear: SLSNe-I are identified by their blue intrinsic colors at low redshift, while at the highest redshifts ($z>1.5$) these events are observationally red.  The next challenges for surveys such as DES, HSC, LSST, \textit{Euclid}, and \textit{WFIRST} is to discover and spectroscopically confirm these objects to $z>3$. 


\acknowledgments
\section*{Acknowledgements}

This paper has gone through internal review by the DES collaboration. It has Fermilab Preprint number PUB$-$17$-$380$-$AE.

We acknowledge support from EU/FP7-ERC grant 615929. R.C.N. would like to acknowledge support from STFC grant ST/N000688/1 and the Faculty of Technology at the University of Portsmouth. L.G. was supported in part by the U.S. National Science Foundation (NSF) under grant AST-1311862. P.J.B.'s work on SLSNe is supported by the {\it Swift} GI program through grant NNX15AR41G. A.V.F.'s group is grateful for financial assistance from the TABASGO Foundation, the Christopher R. Redlich Fund, and the Miller Institute for Basic Research in Science (U.C. Berkeley). We also thank the anonymous referee for comments.
 
Based in part on data obtained from the ESO Science Archive Facility under programs 298.D-5010 and 198.A-0915. 
Based in part on observations obtained at the Gemini Observatory, under program GS-2016B-DD-6, which is operated by the Association of Universities for Research in Astronomy, Inc., under a cooperative agreement with the NSF on behalf of the Gemini partnership: the NSF (United States), the National Research Council (Canada), CONICYT (Chile), Ministerio de Ciencia, Tecnolog\'{i}a e Innovaci\'{o}n Productiva (Argentina), and Minist\'{e}rio da Ci\^{e}ncia, Tecnologia e Inova\c{c}\~{a}o (Brazil).
Based in part on observations made with the NASA/ESA {\it Hubble Space Telescope}, obtained at the Space Telescope Science Institute, which is operated by the Association of Universities for Research in Astronomy, Inc., under NASA contract NAS 5-26555. These observations are associated with program GO-14899. 

Based in part on observations at Cerro Tololo Inter-American Observatory, National Optical Astronomy Observatory, which is operated by the Association of Universities for Research in Astronomy (AURA) under a cooperative agreement with the NSF.
Some of the data presented herein were obtained at the W.M. Keck
Observatory, which is operated as a scientific partnership among the
California Institute of Technology, the University of California, and
NASA; the observatory was made possible by the generous financial
support of the W.M. Keck Foundation. We also acknowledge the help of
T. G. Brink in obtaining the Keck spectrum.

This research used resources of the National Energy Research Scientific Computing Center, a DOE Office of Science User Facility supported by the Office of Science of the U.S. Department of Energy under Contract No. DE-AC02-05CH11231.

Funding for the DES Projects has been provided by the U.S. Department of Energy, the U.S. NSF, the Ministry of Science and Education of Spain, 
the Science and Technology Facilities Council of the United Kingdom, the Higher Education Funding Council for England, the National Center for Supercomputing 
Applications at the University of Illinois at Urbana-Champaign, the Kavli Institute of Cosmological Physics at the University of Chicago, 
the Center for Cosmology and Astro-Particle Physics at the Ohio State University,
the Mitchell Institute for Fundamental Physics and Astronomy at Texas A\&M University, Financiadora de Estudos e Projetos, 
Funda{\c c}{\~a}o Carlos Chagas Filho de Amparo {\`a} Pesquisa do Estado do Rio de Janeiro, Conselho Nacional de Desenvolvimento Cient{\'i}fico e Tecnol{\'o}gico and 
the Minist{\'e}rio da Ci{\^e}ncia, Tecnologia e Inova{\c c}{\~a}o, the Deutsche Forschungsgemeinschaft and the Collaborating Institutions in the Dark Energy Survey. 

The Collaborating Institutions are Argonne National Laboratory, the University of California at Santa Cruz, the University of Cambridge, Centro de Investigaciones Energ{\'e}ticas, 
Medioambientales y Tecnol{\'o}gicas-Madrid, the University of Chicago, University College London, the DES-Brazil Consortium, the University of Edinburgh, 
the Eidgen{\"o}ssische Technische Hochschule (ETH) Z{\"u}rich, 
Fermi National Accelerator Laboratory, the University of Illinois at Urbana-Champaign, the Institut de Ci{\`e}ncies de l'Espai (IEEC/CSIC), 
the Institut de F{\'i}sica d'Altes Energies, Lawrence Berkeley National Laboratory, the Ludwig-Maximilians Universit{\"a}t M{\"u}nchen and the associated Excellence Cluster Universe, 
the University of Michigan, the National Optical Astronomy Observatory, the University of Nottingham, The Ohio State University, the University of Pennsylvania, the University of Portsmouth, 
SLAC National Accelerator Laboratory, Stanford University, the University of Sussex, Texas A\&M University, and the OzDES Membership Consortium.

The DES data management system is supported by the U.S. NSF under grants AST-1138766 and AST-1536171.
The DES participants from Spanish institutions are partially supported by MINECO under grants AYA2015-71825, ESP2015-66861, FPA2015-68048, SEV-2016-0588, SEV-2016-0597, and MDM-2015-0509, 
some of which include ERDF funds from the European Union. IFAE is partially funded by the CERCA program of the Generalitat de Catalunya.
Research leading to these results has received funding from the European Research
Council under the European Union's Seventh Framework Program (FP7/2007-2013) including ERC grant agreements 240672, 291329, and 306478.
We  acknowledge support from the Australian Research Council Centre of Excellence for All-sky Astrophysics (CAASTRO), through project number CE110001020.

This manuscript has been authored by Fermi Research Alliance, LLC under Contract No. DE-AC02-07CH11359 with the U.S. Department of Energy, Office of Science, Office of High Energy Physics. The publisher, by accepting the article for publication, acknowledges that the United States Government retains a non-exclusive, paid-up, irrevocable, world-wide license to publish or reproduce the published form of this manuscript, or allow others to do so, for United States Government purposes.

\facilities{Blanco (DECAM), VLT:Kueyen (X-SHOOTER), Keck:II (DEIMOS), Magellan:Clay (LDSS-3), Gemini:South (Flamingoes-2), HST (WFC3), VLT:Antu (FORS2), VLT:Antu (HAWK-I)} 

\bibliographystyle{apj}
\bibliography{ms}

\begin{deluxetable*}{lllllll}
\tabletypesize{\footnotesize}
\tablecolumns{7}
\tablewidth{0pt}
\tablecaption{DES and SUDSS Light-Curve Data for \nm\label{tab:lc_table}}
\tablehead{
    \colhead{UTC Date} &
    \colhead{MJD} &
    \colhead{Phase} &
    \colhead{$f_g$} &
    \colhead{$f_r$} &
    \colhead{$f_i$} &
    \colhead{$f_z$}\\
    \colhead{} &
    \colhead{} &
    \colhead{(days)\tablenotemark{a}} &
    \colhead{(counts)\tablenotemark{b}} &
    \colhead{(counts)} &
    \colhead{(counts)} &
    \colhead{(counts)}
  }
\startdata
14-01-2016 & $57401.1$ & $-79.4$ &  $-29.3 \pm 59.1$ & $77.3 \pm 72.7$ & $70.5 \pm 108.4$ & $-273.7 \pm 188.6$ \\
21-01-2016 & $57408.1$ & $-77.1$ &  $-16.1 \pm 273.5$ & $175.3 \pm 211.0$ & --- & $232.4 \pm 282.0$ \\
08-02-2016 & $57426.1$ & $-71.1$ &  $-37.2 \pm 145.6$ & $35.2 \pm 188.2$ & $18.5 \pm 267.5$ & $-324.8 \pm 286.6$ \\
09-03-2016 & $57456.1$ & $-61.1$ &  $579.8 \pm 175.0$ & $590.9 \pm 222.0$ & $-225.1 \pm 337.5$ & $749.2 \pm 660.6$ \\
23-08-2016 & $57623.4$ & $-5.3$ &  $-137.5 \pm 281.4$ & $754.3 \pm 199.2$ & $1376.7 \pm 207.6$ & $2305.2 \pm 293.5$ \\
27-08-2016 & $57627.4$ & $-3.9$ &  $352.5 \pm 252.4$ & $1780.2 \pm 344.0$ & $1549.2 \pm 233.8$ & $2960.2 \pm 399.1$ \\
02-09-2016 & $57633.3$ & $-2.0$ &  $318.6 \pm 84.0$ & $1072.9 \pm 116.7$ & $1615.5 \pm 160.3$ & $2457.7 \pm 301.4$ \\
06-09-2016 & $57637.3$ & $-0.6$ &  $178.0 \pm 88.1$ & $897.4 \pm 124.1$ & $1515.5 \pm 175.1$ & $2570.3 \pm 294.8$ \\
12-09-2016 & $57643.2$ & $1.3$ &  --- & $1118.1 \pm 322.0$ & $1139.7 \pm 570.1$ & $2371.8 \pm 442.5$ \\
13-09-2016 & $57644.2$ & $1.7$ &  $-682.6 \pm 1298.3$ & $904.4 \pm 519.9$ & $1089.2 \pm 403.6$ & $2515.2 \pm 629.6$ \\
14-09-2016 & $57645.3$ & $2.0$ &  $230.9 \pm 441.0$ & $1407.3 \pm 262.5$ & $1238.2 \pm 224.5$ & $2614.1 \pm 314.8$ \\
22-09-2016 & $57653.4$ & $4.7$ &  $95.5 \pm 156.6$ & $891.5 \pm 129.2$ & --- & $2360.7 \pm 199.6$ \\
29-09-2016 & $57660.3$ & $7.0$ &  $248.2 \pm 68.1$ & $921.4 \pm 93.9$ & $1259.1 \pm 117.9$ & $2476.9 \pm 228.9$ \\
06-10-2016 & $57667.2$ & $9.3$ &  $283.9 \pm 81.4$ & $1006.9 \pm 103.9$ & $1377.1 \pm 139.9$ & $2335.1 \pm 269.5$ \\
12-10-2016 & $57673.2$ & $11.3$ &  --- & --- & $-1620.3 \pm 1805.7$ & $4610.1 \pm 1744.0$ \\
19-10-2016 & $57680.1$ & $13.7$ &  --- & $639.5 \pm 630.0$ & $1521.4 \pm 625.4$ & $2039.0 \pm 959.1$ \\
20-10-2016 & $57681.2$ & $14.0$ &  --- & $1230.4 \pm 523.2$ & $1114.5 \pm 503.4$ & $1932.1 \pm 623.4$ \\
22-10-2016 & $57683.1$ & $14.6$ &  --- & $780.4 \pm 217.9$ & $2019.4 \pm 326.4$ & $2771.3 \pm 610.7$ \\
25-10-2016 & $57686.1$ & $15.6$ &  $86.1 \pm 130.0$ & $1041.0 \pm 187.5$ & $1870.0 \pm 288.6$ & $2421.5 \pm 521.9$ \\
02-11-2016 & $57694.2$ & $18.3$ &  $134.7 \pm 75.0$ & $728.9 \pm 99.6$ & $852.8 \pm 173.9$ & $2448.8 \pm 308.2$ \\
09-11-2016 & $57701.2$ & $20.7$ &  $187.3 \pm 108.2$ & $610.2 \pm 104.9$ & $1229.7 \pm 129.4$ & $2059.4 \pm 227.6$ \\
10-11-2016 & $57702.3$ & $21.1$ &  --- & --- & --- & --- \\
16-11-2016 & $57708.2$ & $23.0$ &  $-250.2 \pm 530.0$ & $236.3 \pm 270.8$ & $1151.1 \pm 241.6$ & $2108.2 \pm 290.4$ \\
17-11-2016 & $57709.1$ & $23.3$ &  $537.4 \pm 298.4$ & $525.7 \pm 223.8$ & $646.4 \pm 200.3$ & $1707.9 \pm 270.1$ \\
22-11-2016 & $57714.3$ & $25.0$ &  $197.3 \pm 148.0$ & $648.6 \pm 164.9$ & $1203.1 \pm 192.5$ & $1325.7 \pm 379.4$ \\
26-11-2016 & $57718.2$ & $26.4$ &  $278.3 \pm 130.0$ & $377.4 \pm 121.0$ & $259.2 \pm 163.6$ & $1729.6 \pm 336.0$ \\
28-11-2016 & $57720.1$ & $27.0$ &  $235.3 \pm 94.8$ & $331.9 \pm 119.6$ & $625.8 \pm 143.2$ & $1589.7 \pm 253.0$ \\
03-12-2016 & $57725.1$ & $28.6$ &  $202.4 \pm 86.2$ & $568.0 \pm 111.0$ & $423.7 \pm 166.5$ & $1180.5 \pm 327.9$ \\
04-12-2016 & $57726.1$ & $29.0$ &  --- & --- & --- & --- \\
08-12-2016 & $57730.2$ & $30.4$ &  $132.5 \pm 196.1$ & $237.9 \pm 177.5$ & $594.7 \pm 247.3$ & $1538.3 \pm 408.0$ \\
18-12-2016 & $57740.1$ & $33.7$ &  $-19.3 \pm 63.2$ & $220.8 \pm 81.9$ & $440.6 \pm 104.1$ & $1033.9 \pm 188.6$ \\
20-12-2016 & $57742.4$ & $34.4$ &  --- & --- & --- & --- \\
25-12-2016 & $57747.1$ & $36.0$ &  $93.5 \pm 64.6$ & $166.0 \pm 81.2$ & $257.2 \pm 123.5$ & $775.5 \pm 235.5$ \\
29-12-2016 & $57751.1$ & $37.3$ &  $142.0 \pm 88.5$ & $312.2 \pm 108.1$ & $331.5 \pm 156.8$ & $775.2 \pm 299.2$ \\
03-01-2017 & $57756.2$ & $39.0$ &  $67.9 \pm 90.5$ & $1192.3 \pm 117.7$ & $293.8 \pm 160.1$ & $830.9 \pm 318.5$ \\
08-01-2017 & $57761.1$ & $40.6$ &  $-351.0 \pm 528.8$ & $-269.6 \pm 303.1$ & $898.5 \pm 314.5$ & $1302.7 \pm 386.1$ \\
17-01-2017 & $57770.2$ & $43.7$ &  $325.8 \pm 159.7$ & $154.6 \pm 134.3$ & $492.4 \pm 166.1$ & $399.4 \pm 276.4$ \\
25-01-2017 & $57778.1$ & $46.3$ &  $22.9 \pm 106.6$ & $144.8 \pm 118.1$ & $217.1 \pm 164.7$ & $721.6 \pm 329.8$ \\
31-01-2017 & $57784.0$ & $48.3$ &  $115.2 \pm 162.4$ & $205.5 \pm 124.1$ & $480.4 \pm 159.9$ & $544.1 \pm 281.3$ \\
04-02-2017 & $57788.1$ & $49.7$ &  --- & --- & --- & --- \\
07-02-2017 & $57791.1$ & $50.7$ &  $-42.9 \pm 303.4$ & $201.5 \pm 201.3$ & $76.2 \pm 193.9$ & $493.6 \pm 316.5$ \\
08-02-2017 & $57792.0$ & $51.0$ &  --- & --- & --- & --- \\
16-02-2017 & $57800.1$ & $53.7$ &  $-18.8 \pm 113.3$ & $28.4 \pm 159.2$ & $-20.0 \pm 194.8$ & $560.0 \pm 339.7$ \\
\enddata
\tablenotetext{a}{Phase is in the rest frame and relative to maximum light in the DES $i$ band (MJD = 57639.2)}
\tablenotetext{b}{Fluxes $f$ in each filter are given in counts. A zeropoint of 31.0 converts counts into AB magnitudes. No correction for Galactic extinction has been applied.}
\end{deluxetable*}

\begin{deluxetable*}{cccccccc}
\tabletypesize{\footnotesize}
\tablecolumns{7}
\tablewidth{0pt}
\tablecaption{Details of the Spectral Data Obtained for \nm\label{tab:spec_table}}
\tablehead{
    \colhead{UTC Date} &
    \colhead{MJD} &
    \colhead{Phase} &
    \colhead{Telescope} &
    \colhead{Instrument} &
    \colhead{Observed $\lambda$ Range} &
    \colhead{Exp. time} &
    \colhead{Rest $\lambda$ Range}\\
    \colhead{Date} &
    \colhead{} &
    \colhead{(days)\tablenotemark{a}} &
    \colhead{} &
    \colhead{} &
    \colhead{(\AA)} &
    \colhead{(s)} &
    \colhead{(\AA)}
  }
\startdata
09-10-2016 & $57670.0$ & 10.3 & Magellan & LDSS3 & 4500-10000 & 1800 & 2250-5000\\
24-10-2016 & $57685.0$ & 14.9 & VLT & XSHOOTER & 3500-9800 & 9900 & 1750-4900\\
25-10-2016 & $57686.0$ & 15.3 & Keck-II & DEIMOS & 4900-9600 & 6000 & 2250-4800\\
22-11-2016 & $57714.0$ & 24.6 & VLT & XSHOOTER & 3500-9800 & 6300 & 1750-4900\\ 
20-12-2016 & $57742.4$ & 34.4 & HST & WFC3/G141 & 11000-17000 & 2400 & 5500-8500\\
\enddata
\tablenotetext{a}{Phase is in the rest frame and relative to maximum light in the DES $i$ band (MJD = 57639.2).}
\end{deluxetable*}

\begin{deluxetable*}{llllllllll}
\tabletypesize{\footnotesize}
\tablecolumns{7}
\tablewidth{0pt}
\tablecaption{Ancilliary Light-Curve Data for \nm\label{tab:lc_ancilliary_table}}
\tablehead{
    \colhead{UTC Date} &
    \colhead{MJD} &
    \colhead{Phase} &
    \colhead{$f_B$} &
    \colhead{$f_{\text{F2-Y}}$} &
    \colhead{$f_{\text{F2-J}}$} &
    \colhead{$f_{F160W}$} & 
    \colhead{$f_{\text{HAWKI-Y}}$} & 
    \colhead{$f_{\text{HAWKI-J}}$} & 
    \colhead{$f_{\text{HAWKI-H}}$} \\
    \colhead{} &
    \colhead{} &
    \colhead{(days)\tablenotemark{a}} &
    \colhead{(counts)\tablenotemark{b}} &
    \colhead{(counts)} &
    \colhead{(counts)} &
    \colhead{(counts)} &
    \colhead{(counts)} &
    \colhead{(counts)} &
    \colhead{(counts)}
}
\startdata
09-11-2016 & $57701.2$ & $20.7$ &  --- & $2127.8 \pm 166.0$ & $2439.5 \pm 223.2$ & --- & --- & --- & --- \\
10-11-2016 & $57702.3$ & $21.1$ &  $269.2 \pm 23.2$ & --- & --- & --- & --- & --- & --- \\
04-12-2016 & $57726.1$ & $29.0$ &  $209.5 \pm 21.2$ & --- & --- & --- & --- & --- & --- \\
20-12-2016 & $57742.4$ & $34.4$ &  --- & --- & --- & $2255.9 \pm 65.3$ & --- & --- & --- \\
25-12-2016 & $57747.1$ & $36.0$ &  $185.5 \pm 15.6$ & --- & --- & --- & --- & --- & --- \\
03-01-2017 & $57756.2$ & $39.0$ &  --- & $1307.8 \pm 100.9$ & $1725.2 \pm 157.9$ & --- & --- & --- & --- \\
04-02-2017 & $57788.1$ & $49.7$ &  --- & --- & --- & --- & $1027.8 \pm 87.6$ & $1379.5 \pm 117.2$ & $878.5 \pm 184.7$ \\
08-02-2017 & $57792.0$ & $51.0$ &  --- & $1053.9 \pm 112.8$ & $1641.1 \pm 203.2$ & --- & --- & --- & --- \\
\enddata
\tablenotetext{a}{Phase is in the rest frame and relative to maximum light in the DES $i$ band (MJD = 57639.2).}
\tablenotetext{b}{Fluxes $f$ in each filter are given in counts. A zeropoint of 31.0 converts counts into AB magnitudes. No correction for Galactic extinction has been applied.}
\end{deluxetable*}

\end{document}